\documentclass[useAMS,usenatbib]{mn2e}
\usepackage{amsmath}
\usepackage{amssymb}
\usepackage{mathrsfs}
\usepackage{natbib}
\usepackage[english]{babel}
\usepackage{amsfonts}
\usepackage{textcomp}
\usepackage{aas_macros}
\usepackage{times}
\usepackage{graphicx}
\usepackage[usenames,dvipsnames,svgnames,table]{xcolor}
\usepackage[utf8]{inputenc}
\usepackage[T1]{fontenc}
\usepackage{booktabs}

\newcommand{\avg}[1]{\langle{#1}\rangle} 
\newcommand{\sh}{\textrm{sh}\:}
\newcommand{\arcsh}{\textrm{arcsh}\:}
\newcommand{\arctah}{\textrm{arctanh}\:}
\newcommand{\arctanh}{\textrm{arctanh}\:}
\renewcommand{\arctan}{\textrm{arctan}\:}
\newcommand{\tgh}{\textrm{th}\:}

\newcommand{\tah}{\textrm{th}\:}
\newcommand{\Mstar}{M_*}
\newcommand{\rhostar}{\rho_*}
\newcommand{\vh}{v_{\rm h}}
\newcommand{\Phistar}{\Phi_*}
\newcommand{\Phih}{\Phi_{\rm h}}
\newcommand{\Phit}{\Phi_{\rm t}}
\newcommand{\AD}{{\rm AD}}
\newcommand{\Rh}{R_{\rm h}}
\newcommand{\Rc}{R_{\rm c}}
\newcommand{\vphi}{v_{\varphi}}
\newcommand{\vphib}{\overline{v}_{\varphi}}
\newcommand{\vz}{v_z}
\newcommand{\vin}{v_{\rm in}}
\newcommand{\vcirc}{v_{\rm circ}}
\newcommand{\vR}{v_R}
\newcommand{\Ia}{I_{\alpha}}
\newcommand{\Ib}{I_{\beta}}
\newcommand{\Ic}{I_{\gamma}}

\newcommand*{\Msun}{M_{\odot}}
\newcommand*{\kms}{\mathrm{km~s}^{-1}}

\title[Disc galaxies in the Binney potential]{Miyamoto-Nagai discs embedded in the Binney logarithmic potential:
analytical solution of the two-integrals Jeans equations}

\author[C.~O. Smet, S. Posacki \& L. Ciotti]{Christophe Olivier Smet, Silvia Posacki \& Luca Ciotti 
\\
Department of Physics and Astronomy, University of Bologna, viale Berti Pichat 6/2, {\rm 40127} Bologna, Italy}

\date{Accepted, 27 January 2015}

\pagerange{\pageref{firstpage}--\pageref{lastpage}} \pubyear{2014}

\begin{document}
\maketitle
\label{firstpage}

\begin{abstract}

We present the analytical solution of the two-integrals Jeans equations for Miyamoto-Nagai discs embedded in Binney
logarithmic dark matter haloes. The equations can be solved (both with standard methods and with the Residue Theorem)
for arbitrary choices of the parameters, thus providing a very flexible two-component galaxy model, ranging from
flattened discs to spherical systems. A particularly interesting case is obtained when the dark matter halo reduces to
the Singular Isothermal Sphere. Azimuthal motions are separated in the ordered and velocity dispersion components by
using the Satoh decomposition. The obtained formulae can be used in numerical simulations of galactic gas flows, for
testing codes of stellar dynamics, and to study the dependence of the stellar velocity dispersion and of the asymmetric
drift in the equatorial plane as a function of disc and halo flattenings. Here, we estimate the inflow radial velocities
of the interstellar medium, expected by the mixing of the stellar mass losses of the lagging stars in the disc with a
pre-existing gas in circular orbit. 

\end{abstract}

\begin{keywords}
 methods: analytical -- galaxies: kinematics and dynamics -- galaxies: structure -- galaxies: elliptical
and lenticular, cD 
\end{keywords}

\section{Introduction}

Thanks to their simplicity, spherical galaxy models are widely used in stellar
dynamics (e.g., see \citealt{Binney.Tremaine.1987, Bertin.2000}). In particular, the list of galaxy models for which the
Jeans equations have been solved analytically is quite long, both for one and two-component systems (without attempting
at completeness, see, e.g., \citealt{Plummer.1911, Binney.Mamon.1982, Jaffe.1983, Dejonghe.1984, Dejonghe.1986,
Sarazin.White.1987, Hernquist.1990, Renzini.Ciotti.1993, Dehnen.1993, Tremaine.etal1994, Carollo.etal1995,
Ciotti.etal1996, Zhao.1996, Ciotti.1996, Ciotti.1999, Lokas.Mamon.2001, Ciotti.etal2009, VanHese.etal2009}). Even if a
deeper understanding of the model properties can be obtained only by using a phase-space based approach (see e.g.,
\citealt{Michie.1963, King.1966, Wilson.1975, Bertin.Stiavelli.1984, Trenti.Bertin.2005, Binney.2014,
Williams.etal2014}), the moment approach (i.e., the solution of the Jeans equations) is still preferred in applications,
due to the relatively simple method of solution. Of course, in the Jeans approach there is no guarantee that for a
given model the underlying distribution function is positive, and often some educated guess is needed to impose the
closure relation (usually a prescribed radial profile for the velocity dispersion anisotropy). Fortunately, in some
cases it is possible to recover the underlying
phase-space distribution function and check for its positivity (see, e.g., \citealt{Eddington.1916, Osipkov.1979,
Merritt.1985, Cuddeford.1991, Gerhard.1991, Ciotti.Pellegrini.1992, An.Evans.2006,
Ciotti.Morganti.2009,Ciotti.Morganti.2010a, Ciotti.Morganti.2010b}).

Unsurprisingly, the class of axisymmetric galaxy models for which the Jeans equations have been analytically solved is
by far less populated: some examples are the Miyamoto-Nagai model (hereafter MN, \citealt{Miyamoto.Nagai.1975,
Nagai.Miyamoto.1976}), the \citet{Satoh.1980} disc, the family of \citet{Toomre.1982} tori, the flattened Isochrone
\citep{Evans.etal1990}, the Binney logarithmic halo \citep{deZeeuw.etal1996}, the homeoidally expanded systems
\citep{Ciotti.Bertin.2005}, the Ferrers models (e.g., \citealt{Lanzoni.Ciotti.2003}), some systems obtained with the
complex-shift method \citep{Ciotti.Giampieri.2007}, and the power-law systems \citep{Evans.1994,Evans.deZeeuw.1994}.
More recently, two new disc models have been presented \citep{Evans.Bowden.2014,Evans.Williams.2014}, obtained by some
variation of the MN coordinate transformation. Only a handful of two-component axisymmetric galaxy models with
analytical solution of the Jeans equations are available: we recall here the two-component MN models
(\citealt{Ciotti.Pellegrini.1996}, hereafter CP96, where the virial quantities can be expressed analytically), and the
\citet{Evans.1993} phase-space decomposition of the \citet{Binney.1981} logarithmic halo. We finally recall that the
phase-space distribution function approach has been carried out for discs immersed self-consistently in isothermal DM
haloes \citep{Amorisco.Bertin.2010}.

In this paper we show that, quite surprisingly, the Jeans equations for the MN model embedded in the Binney logarithmic
potential can be solved analytically for general choices of the parameters. This model is not new: for example, orbits
in the total potential (with the addition of a spherical bulge), were computed by \citet{Helmi.2004}, in the study
of galactic streamers.

Among the obvious applications of the present model (in addition to test numerical codes dedicated to the solution of
Jeans equations) is the use in hydrodynamical simulations of gas flows in early type galaxies, where
the stellar velocity fields are major ingredients in the description of the energy and momentum source terms due
to the evolving stellar populations (e.g., \citealt{Posacki.etal2013,Negri.etal2014}; see also
\citealt{Pellegrini.2012} and references therein). Another possible application of the present models is to quantify the
effects of the shape of the stellar and dark matter distributions on the kinematics of stars in disc galaxies, a
quantity that can be used to estimate the dark matter amount near the galactic plane (see, e.g.,
\citealt{King.Gilmore.vanderKruit1990}).
Of course, the possibility of a full analytical treatment imposes, as usual, some limitations on the
applicability of the model to describe in detail real observations: here we just recall the fact that disc galaxies show
an exponential profile steeper at large radii than the power-law behaviour of MN profiles. The Binney halo is instead
somehow more realistic (see, e.g., \citealt{Persic.etal1996}), as it leads to a flat rotation curve in the outer parts,
while in the central regions it has a flat core (for non-zero scale lengths), and so it is not fully appropriate for the
description of cuspy dark matter haloes \citep{Dubinski.Carlberg.1991,Navarro.etal1997}.

The paper is organized as follows. In Section~\ref{sec:mod} we present the models and in Section~\ref{sec:solving} we
set up the associated Jeans equations. Their solution, obtained for arbitrary choices of the model parameters with
standard methods, is given in Section~\ref{sec:Binney}; we also prove that the explicit solution can be obtained by
using the Residue Theorem. In Section~\ref{sec:app} we illustrate the main properties of the solutions, and we give the
asymptotic formulae for the most relevant dynamical properties at small and large radii, and near the equatorial plane.
We conclude the analysis by presenting an estimate, based on the asymmetric drift of a specific model, of the inflow
velocity of radial gas flows in the galactic disc, which should be necessarily associated with stellar mass losses of
the lagging stars. The main results are finally summarized in Section~\ref{sec:conc}, while the Appendix contains
technical details and formulae for special cases.

\section{The models} \label{sec:mod}

The models consist of two density components: a stellar MN disc and a dark matter halo characterized by the Binney
logarithmic potential \citep{Binney.Tremaine.1987}. In particular, the MN potential-density pair is
\begin{equation}
\Phistar(R,z)= -\frac{G\Mstar}{\sqrt{R^2+(a+\zeta)^2}},
\label{eq:phistar}
\end{equation}
\begin{equation}
\rhostar(R,z)= \frac{\Mstar b^2}{4\pi} \frac{aR^2+(a+3\zeta)(a+\zeta)^2}{\zeta^3[R^2+(a+\zeta)^2]^{5/2}},
\label{eq:rhostar}
\end{equation}
where $\zeta=\sqrt{z^2+b^2}$ and $(R,\varphi,z)$ are the standard
cylindrical coordinates. 

The Binney logarithmic family is defined by
\begin{equation}\label{eq:Binneypot}
\Phih(R,z)=\frac{\vh^2}{2}\ln\left(\Rh^2+R^2+\frac{z^2}{q^2}\right),
\end{equation}
where $q$ is the axis ratio of the equipotential surfaces, and $\vh$ and $\Rh$ are constants related to the halo
circular velocity in the equatorial plane as
\begin{equation}
\vcirc (R)=\frac{\vh R}{\sqrt{\Rh^2+R^2}}.
\end{equation}
Note that $\vcirc$ is independent of $q$, and that if $q<1/\sqrt{2}$, the halo density is no longer everywhere positive,
independently of the value of $\Rh$ \citep{Binney.Tremaine.1987}. Moreover, for zero flattening ($q=1$) and $\Rh=0$
we have the special but important case of the Singular Isothermal Sphere (SIS) \footnote{Without loss of generality, in
eqs.~(\ref{eq:Binneypot}) and (\ref{eq:SISpot}) the argument of the logarithm is implicitly assumed normalized to some
scale length.}
\begin{equation}\label{eq:SISpot}
\Phih(r)=\vh^2\ln r,
\end{equation}
where $r=\sqrt{R^2+z^2}$ is the spherical radius, and $\vcirc (R)=\vh$.

Notice that this new two-component model allows a few interesting limiting cases: for example, one can consider a
stellar component with spherical symmetry ($a=0$ in the MN model) within a Binney logarithmic halo which has no
spherical symmetry ($q\neq 1$), or alternatively a non-spherical stellar component ($a>0$) within a spherical halo
($q=1$). We can finally introduce the spherical symmetry in both components, by choosing $a=0$ in the MN model and $q=1$
in the Binney logarithmic halo.

\section{The Jeans equations}\label{sec:solving}

For an axisymmetric distribution $\rhostar (R,z)$ supported by a two-integrals phase-space distribution function
$f(E,J_z)$, the Jeans equations for the stellar component write (e.g. \citealt{Binney.Tremaine.1987})
\begin{equation}
\frac{\partial\rhostar\sigma_*^2}{\partial z}=-\rhostar \frac{\partial\Phit}{\partial z},
\label{eq:J1}
\end{equation}
and
\begin{equation}
\frac{\partial\rhostar\sigma_*^2}{\partial R}-
\rhostar\frac{\overline{\vphi^2}-\sigma_*^2}{R}=-\rhostar\frac{\partial\Phit}{\partial R},
\label{eq:J2}
\end{equation}
where $\Phit =\Phistar +\Phih$. These equations are simplified with respect to the general case because for a
two-integrals system: (1) the velocity dispersion tensor is aligned with the coordinate system, i.e. the phase-space
average of the mixed products of the velocity components vanish,
$\overline{\vR\vz}=\overline{\vR\vphi}=\overline{\vphi\vz}=0$; (2) the radial and vertical velocity dispersions are
equal, i.e. $\sigma_R=\sigma_z\equiv\sigma_*$; (3) the only possible non-zero streaming motion is in the azimuthal
direction. 

Of course, in real galaxies the situation can be more complicated. For example, 
different investigations, carried out within both RAVE (the RAdial Velocity Experiment) and SDSS (the Sloan
Digital Sky Survey), find that the velocity ellipsoid of disc stars in the Milky Way tilts on moving above or below the
Galactic plane, reaching a tilt angle of $7-10$ degrees at a height of 1 kpc \citep{Siebert.etal2008,Smith.etal2012}.
Considering again the Milky Way, \citet{Smith.etal2012} find also a ratio
$\sigma_z/\sigma_R=0.6-0.8$ dependent on metallicity, and there is also some evidence for this ratio to be different
from unity for external disc galaxies (e.g., \citealt{vanderKruit.deGrijs1999}).

Finally, note that in the special case of spherical symmetry, in \textit{both} the stellar component and the dark
matter halo, formal integration of eq.~(\ref{eq:J1}) shows that $\sigma_*$ has spherical symmetry, and from
eq.~(\ref{eq:J2}) $\overline{\vphi^2}=\sigma_*^2$.

In order to split the azimuthal velocity field in its ordered and random components we adopt the \citet{Satoh.1980}
$k$-decomposition
\begin{equation} \label{eq:Satoh}
\vphib^2=k^2(\overline{\vphi^2}-\sigma_*^2),
\end{equation}
so that
\begin{equation} \label{eq:Satohcons}
\sigma_{\varphi}^2\equiv\overline{\vphi^2}-\vphib^2=k^2\sigma_*^2+(1-k^2)\overline{\vphi^2},
\end{equation}
where $0\leq k\leq 1$. This implicitly assumes that the supporting phase-space distribution function depends on $k$,
i.e., $f=f(E,J_z;k)$. Note that, while in the Satoh decomposition $k$ is independent of position, in principle, $k^2$
can be a function of $(R,z)$, bounded above by the function $k_{\rm max}(R,z)$, defined by the condition
$\sigma^2_{\varphi}=0$ everywhere (CP96). The case $k=1$ corresponds to the isotropic rotator while for $k=0$ no net
rotation is present and all the flattening of $\rhostar$ is due to the azimuthal velocity dispersion $\sigma_{\varphi}$.
Note finally that if the distribution function is of the Satoh family, then the spherical limit is isotropic and
non-rotating independently of $k$, as can be seen from eq.~(\ref{eq:Satohcons}) with $\overline{\vphi^2}=\sigma_*^2$.
Clearly, the possibility of using the Satoh decomposition depends on the positivity of the r.h.s. of
eq.~(\ref{eq:Satoh}), a condition that can be violated for arbitrary choices of the density components of the model.
This problem is related to the analogous issue encountered in the construction of rotating baroclinic configurations
of assigned density distribution in Fluidodynamics (where of course eqs.~\ref{eq:J1} and \ref{eq:J2} are restricted to
the isotropic case, and the velocity dispersion is substituted by the thermodynamic pressure). We will return on this
point in Section~\ref{sec:radial}.

In the following, we will use the parameter $s=a/b$ to quantify the flattening of the MN
disc. The choice $a=0$ ($s=0$) gives the \citet{Plummer.1911} sphere, while $b=0$ ($s\to\infty$) gives the
razor-thin \citet{Kuzmin.1956} disc. As we do not consider this case of the Kuzmin disc\footnote{Formally, a razor-thin
disc supported by a two-integrals phase-space distribution can have only circular orbits.}, in the following all the
lengths will be normalized to $b$ (in order to avoid cumbersome notation, from now on $R$, $\Rh$ and $z$ must be
intended normalized to $b$, if not differently stated).

The Jeans equations have been solved analytically for the one-component model (\citealt{Miyamoto.Nagai.1975}; see also
CP96, where analytical expressions for the virial quantities are derived for two-component MN models with different
flattenings of the two components, but the same scale $b$). Easy algebra shows that
\begin{equation}
\label{eq:jeans1solnohalo}
\rhostar\sigma_{**}^2 =\frac{G\Mstar^2}{8\pi b^4}\frac{(s+\zeta)^2}{\zeta^2[R^2+(s+\zeta)^2]^3},
\end{equation}
and
\begin{equation}
\label{eq:jeans2solnohalo}
\rhostar(\overline{\vphi^2}-\sigma_{**}^2)=\frac{G\Mstar^2}{4\pi
 b^4}\frac{s R^2}{\zeta^3[R^2+(s+\zeta)^2]^3},
\end{equation}
where now $\zeta=\sqrt{1+z^2}$. The subscript "$_{**}$" indicates a quantity associated
with the self-interaction of the stellar component, while "$_{\rm *h}$" indicates the terms due to the effect of the
dark matter halo on the stellar component; in particular, in eqs.~(\ref{eq:J1}) and (\ref{eq:J2})
$\sigma_*^2=\sigma^2_{**}+\sigma^2_{\rm  *h}$.
An important global quantity of the model is the virial interaction energy,
$W=-\int\rhostar\avg{\mathbf{x},\nabla\Phit}d^3\mathbf{x}=W_{**}+W_{\rm *h}$. For the MN model (CP96), 
\begin{equation}
\label{eq:W**}
W_{**}=\dfrac{G\Mstar^2}{8b}\left[\dfrac{\pi}{2s^2}-\dfrac{1-2s^2}{s(1-s^2)}-\dfrac{F(s)}{s^2(1-s^2)} \right],
\end{equation}
where\footnote{For a typo, the fraction marks in eq.~(A1) in CP96 are missing.}
\begin{equation}
F(s)=
\begin{cases}
\displaystyle\arccos(s)/\sqrt{1-s^2}, &0\leqslant s < 1,\\
\displaystyle 1+(1-s)/3+\mathcal{O}(1-s)^2, &s\to 1,\\
\displaystyle\textrm{arccosh}(s)/\sqrt{s^2-1}, &s>1,
\end{cases}
\end{equation}
and so 
\begin{equation} 
W_{**}(0)=-\frac{3\pi G\Mstar^2}{32b},
\end{equation} 
\begin{equation} 
W_{**}(1)=-\left(\frac{1}{3}-\frac{\pi}{16}\right)\dfrac{G\Mstar^2}{b}.
\end{equation}
In the present models
\begin{equation}
W_{\rm *h}=-\vh^2\int\rhostar\dfrac{q^2R^2+z^2}{q^2R^2+z^2+q^2\Rh^2}\,d^3\mathbf{x},
\label{eq:WBin}
\end{equation} 
so that $W_{\rm *h}$ increases for increasing $\Rh$.
In general, for $\Rh>0$ the integral above cannot be expressed in terms of elementary functions (with the trivial
exception of
spherical symmetry of the two components). Remarkably, eq.~(\ref{eq:WBin}) shows that in the case of a coreless
logarithmic potential ($\Rh=0$), $W_{\rm *h}=-\vh^2\Mstar$, for arbitrary $\rho_*$ of finite total mass $M_*$.

\section{The solution}\label{sec:Binney}
\subsection{The vertical Jeans equation}\label{subsec:vert}

In the case of a Binney logarithmic halo, the halo contribution to the vertical and radial velocity dispersion is given
by
\begin{equation}
\rhostar\sigma_{\rm *h}^2= \int_z^\infty\rhostar\frac{\partial\Phih}{\partial z'}dz'=
\int_z^\infty \frac{\vh^2\,\rhostar z'\: dz'}{A+1+z'^2}=
\frac{\Mstar\vh^2}{ 4\pi b^3} I,
\label{eq:binney}
\end{equation}
where the last identity is obtained by normalization of all the lengths to $b$, and $I$ is a dimensionless function. In
eq.~(\ref{eq:binney}) $A\equiv q^2(R^2+\Rh^2)-1$. Note that, given $q$ and $\Rh$, the minimum value for the quantity $A$
is $q^2\Rh^2-1$, a value reached on the $z$ axis; for a coreless halo, $A\geq -1$. As we will see, the sign of $A$
plays an important role in the treatment of eq.~(\ref{eq:binney}). In fact, for $q\Rh > 1$, $A$ is positive
independently of $R$, while for $q\Rh <1$ a radius $\Rc\equiv\sqrt{1/q^2 -\Rh^2}$ exists so that for $R <\Rc$ the
parameter $A$ is negative. We call $\Rc$ the \textit{critical radius}, and the surface $R =\Rc$ the \textit{critical
cylinder}. In the special case of the SIS halo, $\Rc=1$. The integral in eq.~(\ref{eq:binney}) is quite formidable,
especially considering the fact that $\rhostar$ contains two nested irrationalities (see eq.~\ref{eq:rhostar}).
Surprisingly, in the following we show that this integral can in fact be computed in terms of elementary functions. We
stress that there are special values of the parameters for which the general treatment described below cannot be used
(or can be significantly simplified), and the corresponding formulae should be obtained from the general ones with some
careful limit process. Even if this approach does not present conceptual difficulties, we prefer to list these special
cases in Table~\ref{tab:special}, referring to Appendix~\ref{app:special}, where the explicit formulae are provided.

In order to integrate eq.~(\ref{eq:binney}) we begin by removing the inner
irrationality in $\rhostar$ with the substitution $\zeta=\sqrt{1+z^2}$, so that
\begin{equation}
I\equiv \int_\zeta^\infty \frac{sR^2+(s+3\zeta')(s+\zeta')^2}{
 \zeta'^2[R^2+(s+\zeta')^2]^{5/2}(A+\zeta'^2)}d\zeta'.
 \label{eq:t-int}
\end{equation}
Note that $\zeta\geq 1$, and so $A+\zeta^2 \geq 0$ everywhere: equality holds at the origin only for a coreless halo,
i.e. for $\Rh=0$ (in particular for the SIS). In order to proceed with the integration, we now remove the second
irrationality with the change of variable $\sh x=(s+\zeta)/R$. This substitution is not valid on the $z$-axis, however
in this case the integrand in eq.~(\ref{eq:t-int}) is a rational function of $\zeta'$ and its integration is elementary
(Table~\ref{tab:special}, first case). Restricting to $R> 0$ we obtain
\begin{equation}
I=\frac{1}{R^2}\int_{\arcsh\lambda}^\infty \frac{\left[s+(3R\,\sh
 x-2s)\sh^2x\right]\: dx}{(R\,\sh x-s)^2[A+(R\,\sh x-s)^2](1+\sh^2 x)^2},
\label{eq:totalint}
\end{equation}
where $\arcsh x=\ln\left(x+\sqrt{1+x^2}\right)$ and 
\begin{equation} 
\lambda\equiv\frac{s+\zeta}{R}=\frac{s+\sqrt{1+z^2}}{R}.
\label{adef}
\end{equation} 

\begin{table}
\centering
\caption{Special cases}
\begin{tabular}{ccc}
\toprule
Name                       &            & Appendix                 \\
\midrule                           
$z$ axis                   & $R=0$      & \ref{sec:app:R0}         \\
critical cylinder          & $R=\Rc$    & \ref{sec:I2A0}           \\
spherical stellar density  & $s=0$      & \ref{sec:app:plummer}    \\
\bottomrule
\end{tabular}
\flushleft
Notes: special cases for which the treatment in Section~\ref{sec:Binney}
is no longer valid or can be greatly simplified. The corresponding formulae
are given in the specific sections of Appendix~\ref{app:special}.
\label{tab:special}
\end{table}

Integral (\ref{eq:totalint}) is our main equation. The idea is to expand the integrand by using the
\citet{Hermite.1872} partial
fraction decomposition in terms of $\sh x$. Following Appendix~\ref{app:intcos} one obtains
\begin{equation}
I=\frac{\Ia+\Ib+\Ic}{ R^2}.
\label{eq:I123}
\end{equation}
Notice however that if $A=0$ (i.e., for $R=\Rc$), two of the denominator factors in eq.~(\ref{eq:totalint}) coincide,
and a different partial fraction decomposition is needed (Table~\ref{tab:special}, second case). In the case of a
spherical stellar distribution, $s=0$ and the integrand in eq.~(\ref{eq:t-int}) also simplifies
(Table~\ref{tab:special}, third case). In conclusion, we now focus on the evaluation of eq.~(\ref{eq:totalint}) in the
case $R\neq 0, A\neq 0, s\neq 0$, i.e. for points not on the $z$-axis, on the critical cylinder, and for non-spherical
stellar densities.

The integral $\Ia$ in eq.~(\ref{eq:I123}) is of trivial evaluation and, following Appendix~\ref{sec:app:Ia}, the result
is
\begin{equation}
\begin{split}
\Ia=&\frac{2\alpha_0+\alpha_2}{3}+\frac{\alpha_3}{\sqrt{1+\lambda^2}}-\frac{\alpha_0\lambda}{(1+\lambda^2)^{3/2}}\\
&-\frac{\alpha_3-\alpha_1}{3(1+\lambda^2)^{3/2}}-\frac{(\alpha_2+2\alpha_0)\lambda^3}{3(1+\lambda^2)^{3/2}}.
\end{split}
\label{eq:Ia}
\end{equation}

For the computation of $\Ib$ we use the standard substitution $y=\tgh(x/2)$ to obtain a rational integrand, which can be
solved again by partial fraction decomposition. With this substitution the upper limit of integration in
eq.~(\ref{eq:totalint}) becomes $1$, while the lower limit of integration becomes
\begin{equation}
\mu\equiv\tanh\left(\frac{\arcsh\lambda}{2}\right)=\dfrac{\sqrt{1+\lambda^2}-1}{\lambda},
\label{eq:y0}
\end{equation} 
and from Appendix~\ref{sec:app:Ib}
\begin{equation}
\Ib=\frac{\beta_0}{R}\left(\frac{\sqrt{1+\lambda^2}}{\zeta}-\frac{1}{R}\right).
\label{eq:Ibsol}
\end{equation}

The integral $\Ic$ is the most difficult one, due to the presence of the parameter $A$ in the integrand. The
exponential substitution $y=e^x$ is now adopted, leading to a rational integrand. The upper limit of
integration becomes $\infty$, while the lower limit of integration becomes
\begin{equation}
\nu\equiv e^{\arcsh\lambda}=\lambda+\sqrt{1+\lambda^2}.
\label{eq:nu}
\end{equation}
All the details about the integration are given in Appendix~\ref{sec:app:Ic}, and the final expression is\footnote{The
addition formulae used in eq.~(\ref{eq:Icsol}) are $\arctanh u-\arctanh v= \arctanh \frac{u-v}{1-uv}$ and $\arctan
u-\arctan v = \arctan \frac{u-v}{1+uv}$, where for $|x|<1$, $\arctah x= \frac{1}{2}\ln\frac{1+x}{1-x}.$}
\begin{equation}
\begin{split}
\Ic=&\frac{\gamma_1\eta_+}{R^2}\ln\frac{(\nu-\Delta_-)^2+\delta_-}{(\nu-\Delta_+)^2+\delta_+}+\frac{2\gamma_1}{R^2}\frac
{\theta_++\eta_+\Delta_+}{\sqrt{|
\delta_+|}}\times\\
&\begin{cases}
\displaystyle\arctan\frac{\sqrt{\delta_+}(\nu-\Delta_-)-\sqrt{\delta_-}(\nu-\Delta_+)}{
(\nu-\Delta_+)(\nu-\Delta_-)+\sqrt{\delta_+\delta_-}},&A>0,\\
\displaystyle\arctanh\frac{\sqrt{|\delta_+|}(\nu-\Delta_-)-\sqrt{|\delta_-|}(\nu-\Delta_+)}{
(\nu-\Delta_+)(\nu-\Delta_-)-\sqrt{\delta_+\delta_-}},&A<0.
\end{cases}
\end{split}
\label{eq:Icsol}
\end{equation}

The explicit expressions for the constants appearing in eqs.~(\ref{eq:Ia}), (\ref{eq:Ibsol}) and (\ref{eq:Icsol}) are
given in Appendix~\ref{app:intcos}. We stress again that the formulae above cannot be used in their present form to
describe the velocity dispersion on the $z$ axis ($R=0$), on the critical cylinder ($A=0$, i.e., $R=\Rc$), or in the
case of a spherical stellar density ($s=0$). As listed in Table~\ref{tab:special}, all these cases are treated
separately in Appendix~\ref{app:special}.

\subsection{The radial Jeans equation}
\label{sec:radial}
In the previous Section we solved the vertical Jeans eq.~(\ref{eq:J1}). For the radial Jeans eq.~(\ref{eq:J2}), no
further integration is needed, as only the radial derivative of $I$ is required to compute $\overline{\vphi^2}
-\sigma^2_*$. In principle, one can obtain an explicit expression for $dI/dR$ by differentiating eq.~(\ref{eq:I123}).
However, the easiest way to obtain $dI/dR$ is to differentiate eq.~(\ref{eq:totalint}) with respect to $R$, and to
perform the partial fraction decomposition on the resulting integrand. The resulting integrals are formally similar to
the ones already computed, and they can be solved with the same techniques. In any case, we chose not to include these
explicit expressions here, since available computer algebra systems can easily perform the differentiation of
eq.~(\ref{eq:I123}).

\begin{figure}
\includegraphics[width=\linewidth]{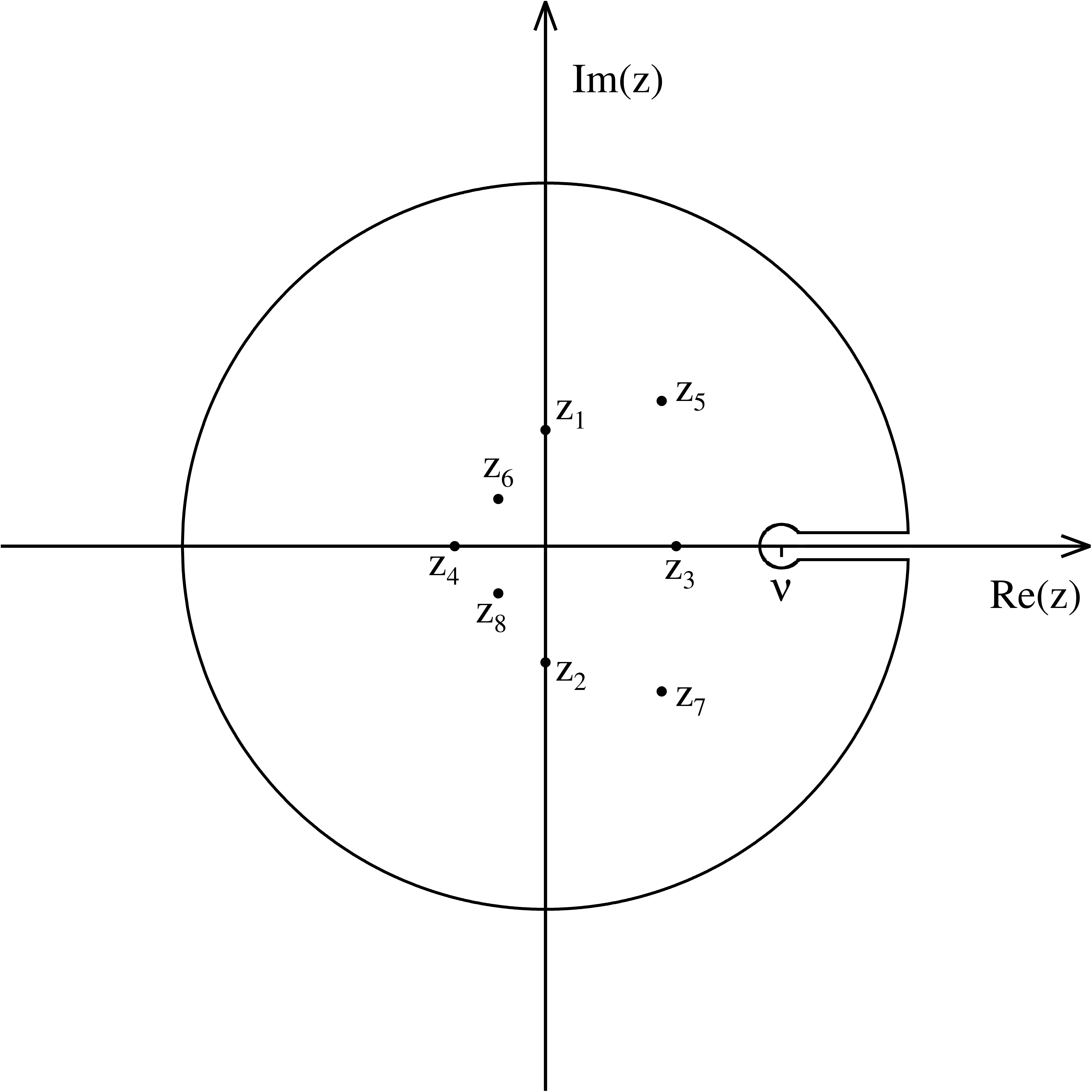}
\caption{The integration contour with the poles for the function $f_0(y)\ln(y-\nu)$ introduced in
Section~\ref{sec:complex}.
The figure corresponds to points $(R,z)$ outside the critical cylinder, i.e. $A>0$ in eq.~(\ref{eq:totalint}).}
\label{fig:contour}
\end{figure}
For general discussions, it is useful to provide a very elegant commutator-like formula for the quantity
$\overline{v^2_{\varphi}}-\sigma_*^2$, related to the study of
rotating gaseous baroclinic equilibria (see \citealt{Barnabe.etal2006}, and references
therein):
\begin{equation}
\overline{\vphi^2} -\sigma_*^2 =\dfrac{R}{\rho_*}\int_z^{\infty}\left(\dfrac{\partial \rho_*}{\partial
R} \dfrac{\partial
\Phit}{\partial z'}-\dfrac{\partial \rho_*}{\partial z'} \dfrac{\partial \Phit}{\partial R} \right)dz' .
\label{eq:commutator}
\end{equation}
It is not surprising that this relation appears both in Fluidodynamics \citep{Rosseland.1926,Waxman.1978} and in Stellar
Dynamics \citep{Hunter.1977}, due to the strict relation of the Jeans and fluid equations. We will use
eq.~(\ref{eq:commutator}) for some consideration about the sign of $\overline{v^2_{\varphi}}-\sigma_*^2$ in
Section~\ref{sec:app}, while here we just note how from eq.~(\ref{eq:commutator}) it follows that for fully spherical
models (stars plus dark matter) the commutator vanishes identically, so that, in the Satoh decomposition, such spherical
models cannot rotate and are isotropic, a conclusion already reached by using other arguments in
Section~\ref{sec:solving}. Finally, in the case of a spherical stellar density, the only non-zero contribution to
eq.~(\ref{eq:commutator}) (and so to rotation in the Satoh decomposition, see eq.~\ref{eq:Satoh}) can be produced by a
non-spherical dark matter halo.

\subsection[]{The contour integral approach} \label{sec:complex}

We conclude this Section by noticing that, quite remarkably, the integral in eq.~(\ref{eq:totalint}) can be expressed in
explicit form also by using the Residue Theorem of Complex Analysis. With the change of variable $y=e^x$ in
eq.~(\ref{eq:totalint}) we obtain a rational integrand in $y$, which we call $f_0(y)$, the integration domain
becomes $\nu\leqslant y <\infty$, and $\nu$ is given by eq.~(\ref{eq:nu}). It is trivial to prove that the denominator
of $f_0(y)$ has no real zeros in this range. The numerator of $f_0$ has degree 11 and the denominator has degree 16, so
that, when considering $y$ a complex variable, the integrand amply satisfies Jordan's Lemma (e.g.,
\citealt{Titchmarsh.1932}). The Residue Theorem cannot be applied directly to $f_0$ however, as after one circuitation
around the origin in the complex plane, the two contributions on the real axis would cancel exactly. This problem can be
obviated by applying the Residue Theorem to the suitably modified function $f_0(y)\ln(y-\nu)$, where the integration
path is schematically
given in Fig.~\ref{fig:contour}, so that
\begin{equation}
I=\int_{\nu}^{\infty}f_0(y)dy= - \sum {\rm Res}[f_0(y)\ln(y-\nu)].
\end{equation}
The result holds because the line integrals along the large and small circular paths vanish in the limits of infinite
large and small radii, while the $2\pi i$ additive constant, produced by the multivalued factor $\ln(y-\nu)$ along the
real axis after the circuitation, leads to the final result. In Fig.~\ref{fig:contour} we show, for illustrative
purposes, the positions of the poles of
$f_0$ for a generic model, and for $R$ outside the critical cylinder ($A >0$). As the coefficients appearing in $f_0$
are real, the poles are either real or they come in complex conjugate pairs; moreover, they are both fixed and movable
(i.e., dependent on $R$). Remarkably, all these poles can be expressed as simple algebraic functions, so that the
residues can be also calculated explicitly. In fact, there are two fixed {\it quadruple} poles at $z_{1,2}=\pm i$, two
movable {\it double} real poles at $z_{3,4}=(s\pm\sqrt{s^2+R^2})/R$, and four movable {\it simple} poles at
$z_{5,6}=(u\pm\sqrt{u^2+R^2})/R$ and $z_{7,8}=(v\pm\sqrt{v^2+R^2})/R^2$, where $u=s+\sqrt{-A}$ and $v=s-\sqrt{-A}$.
These last four poles are real inside the critical cylinder ($A<0$), while they come in two pairs of complex conjugate
simple poles outside the critical cylinder; finally on the critical cylinder they merge with the other double real poles
$z_{3,4}$, yielding two quadruple real poles. Overall, despite the elegance of this method and the interesting property
that the integration can be performed in closed form by using the Residue Theorem, the number of poles, their
multiplicity, and their complex nature do not reduce the amount of work needed to obtain the final (real) result when
compared with the standard method in Section~\ref{subsec:vert}.

\begin{figure*}
\includegraphics[width=0.327\linewidth]{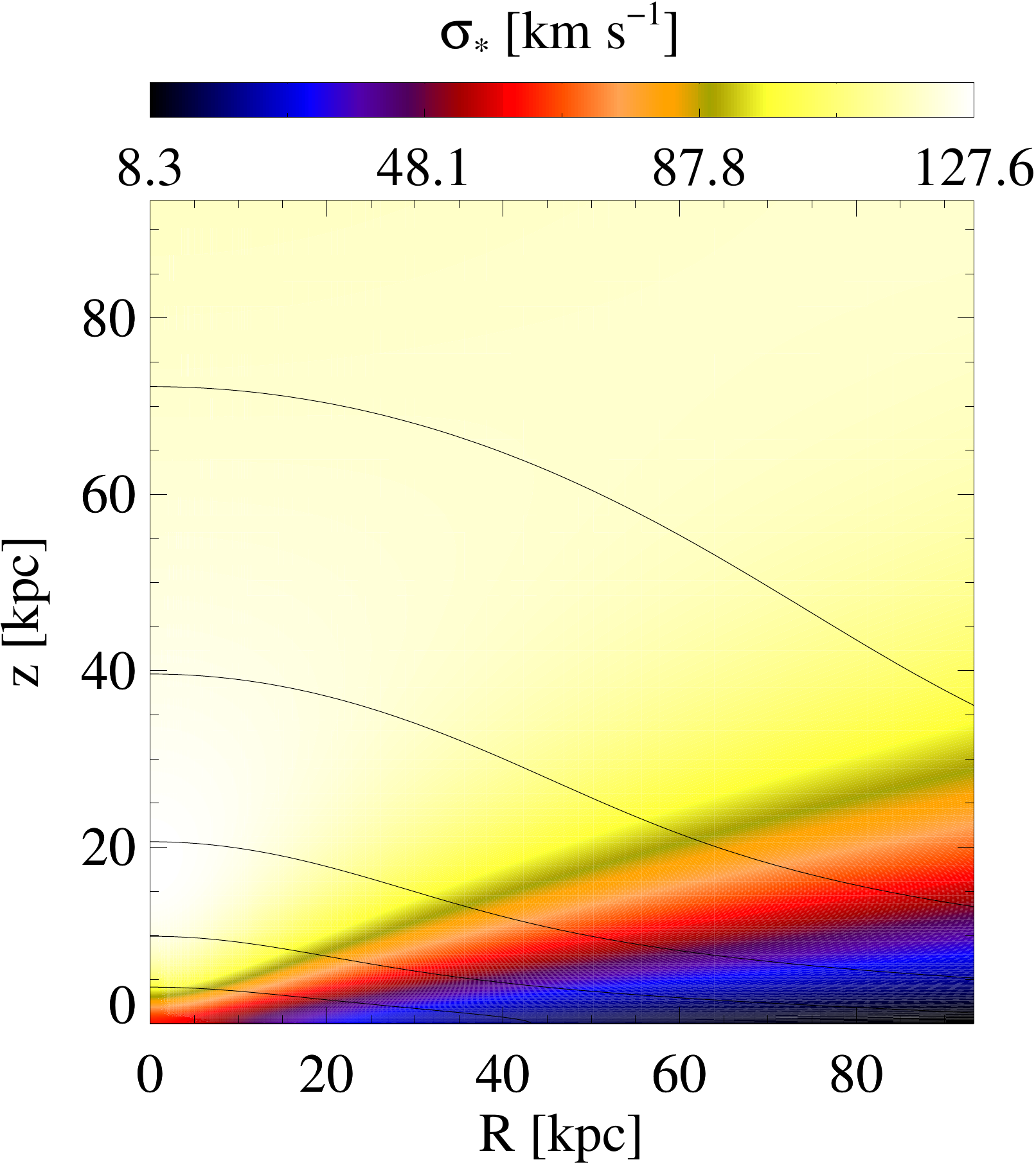}\hspace{2mm}
\includegraphics[width=0.32\linewidth]{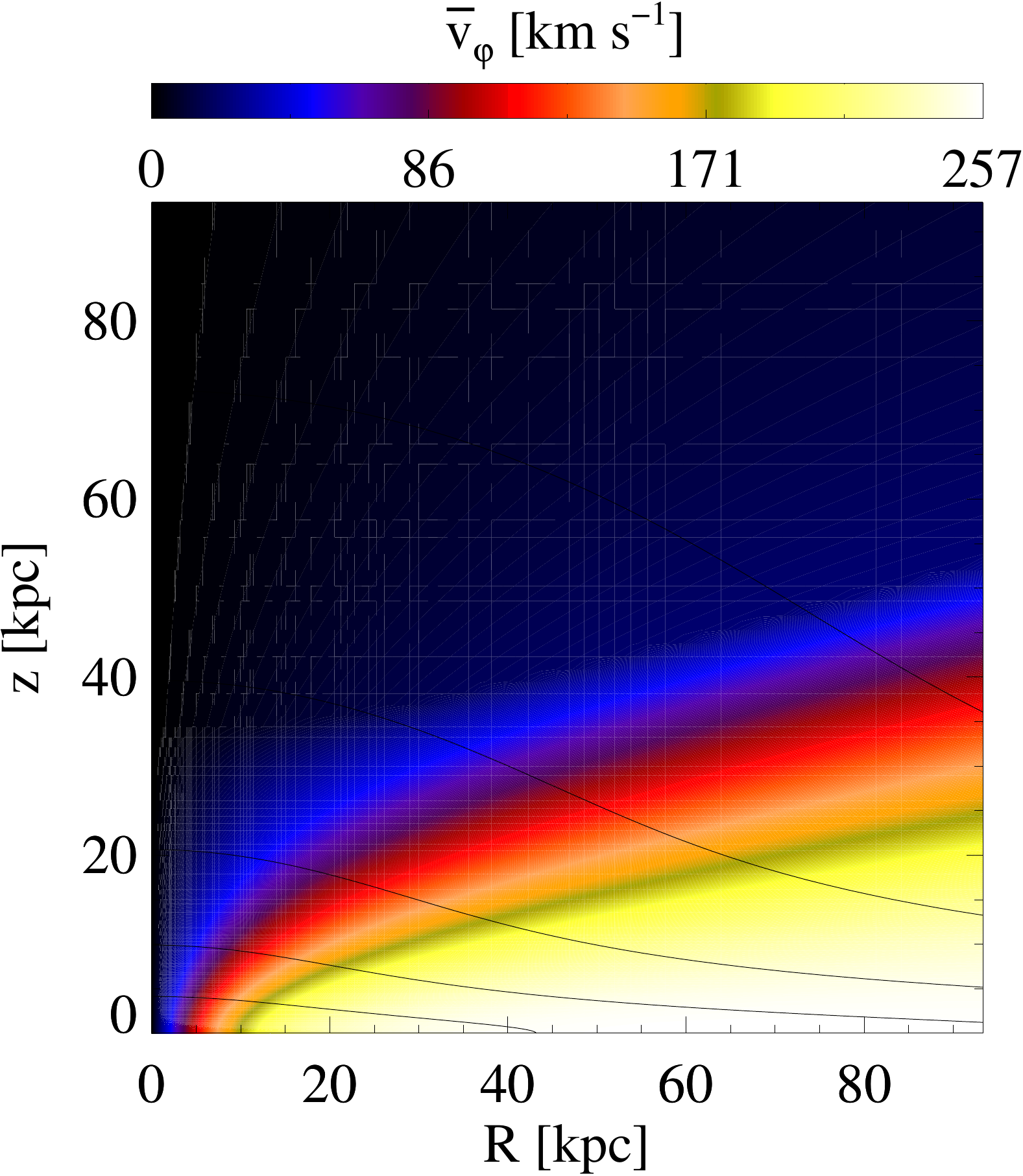}\hspace{2mm}
\includegraphics[width=0.32\linewidth]{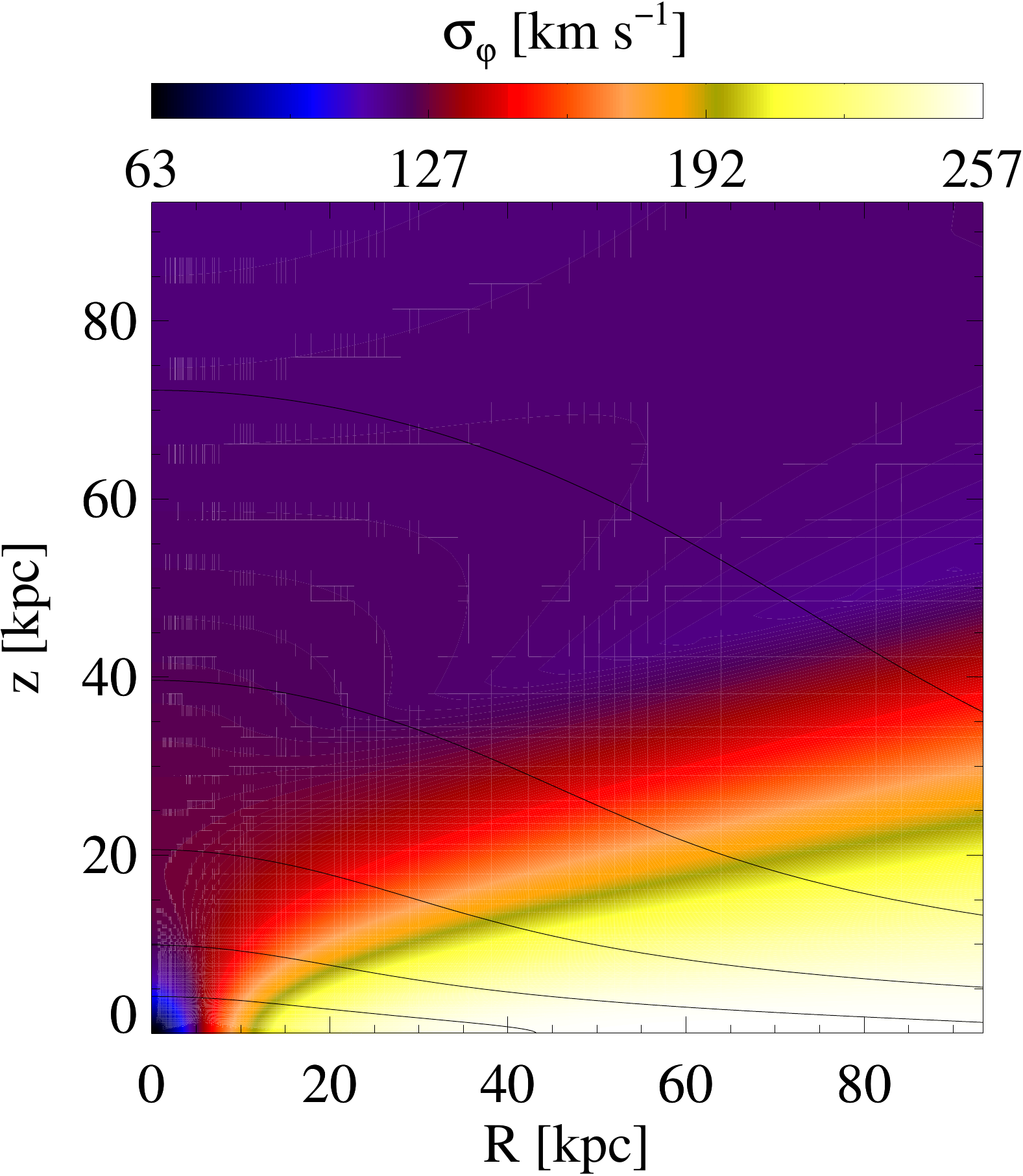}
\caption{Two-dimensional maps in the meridional plane of the vertical and radial velocity dispersion
$\sigma_*=\sqrt{\sigma^2_{**} + \sigma^2_{\rm *h}}$ (left panel), of the ordered azimuthal velocity $\vphib$ in the
isotropic case ($k=1$, central panel), and of the azimuthal velocity dispersion $\sigma_{\varphi}$ in the fully velocity
dispersion supported case ($k=0$, right panel). The structural parameters of the model are $\Mstar=10^{11} \Msun$, $b=2$
kpc, $s=10$, $\vh=250\,\kms$, $\Rh=5b$, and $q=0.7$. Solid lines represent isodensity contours of the stellar
distribution.}
\label{fig:2Dmap}
\end{figure*}
\section{General properties of the solution}\label{sec:app}

As the Jeans equations of the one component MN model have been already solved (eq.~\ref{eq:jeans1solnohalo} and
\ref{eq:jeans2solnohalo}), from the results of Section~\ref{sec:Binney} it follows that a new two-component
axisymmetric galaxy model, admitting a fully analytical solution for the Jeans equations for the stellar component is
now available.
The formulae reported in the text (and in Appendix~\ref{app:special}) are fully general and they can
be easily implemented in numerical codes and in computer algebra systems to explore the behaviour of the kinematical
fields of the model in all cases of interest. However, the obtained formulae are sufficiently cumbersome to prevent an
immediate reading of their physical content, so that a graphical exploration is useful for a qualitative understanding.

A first illustration of the behaviour of the solutions in the meridional $(R,z)$ plane is given in
Fig.~\ref{fig:2Dmap}, where we show the fields $\sigma_*$ (left panel), $\vphib$ (central panel) and $\sigma_{\varphi}$
(right panel), for a quite flat MN stellar distribution ($s=10$) of total mass $M_*=10^{11} \Msun$, and
scale-length $b=2$ kpc, embedded in a maximally flattened DM logarithmic halo with $\Rh=10$ kpc, $\vh=250\, \kms$ and
$q=0.7$. For reference, the solid lines show the isodensities of the stellar distribution. As expected, the vertical
velocity dispersion $\sigma_*$ near the
equatorial plane (independent of the specific decomposition of the azimuthal fields) declines for increasing $R$, due
to the
fact that the MN stellar distribution becomes more and more flat. Of course, in the isotropic case
$\sigma_{\varphi}=\sigma_*$, and the flattening of the stellar distribution is supported by ordered rotation (central
panel). In the fully velocity dispersion supported model (right panel), while $\sigma_*$ is the same as in the left
panel, $\sigma_{\varphi}$ now supports the flattening, with a field very similar to $\vphib$ of the isotropic rotator.
The figures can be compared with the analogous plots in Negri et al. (2014, Fig.~1), which were relative to a rounder
stellar MN model ($s=1$), embedded in an Einasto dark matter halo (of exponent $n=4$). Although the models are not
exactly the same, the overall structure of the kinematical fields, both in the isotropic ($k=1$) and fully velocity
dispersion supported ($k=0$) cases are very similar. The only significant difference is that of an 'hourglass' structure
in $\sigma_*$ that can be observed near the centre of the \citet{Negri.etal2014} models, but is missing in
Fig.~\ref{fig:2Dmap}. The hourglass shaped distribution is a characteristic of $\sigma_{**}$ for the MN model, and in
the present models is absent because of the contribution of the relatively massive Binney halo adopted (at variance with
the lighter Einasto halo in \citealt{Negri.etal2014}): in fact, we checked that, for decreasing values of $\vh$, this
feature becomes again evident in the present models.

\begin{figure*}
\includegraphics[width=0.32\linewidth]{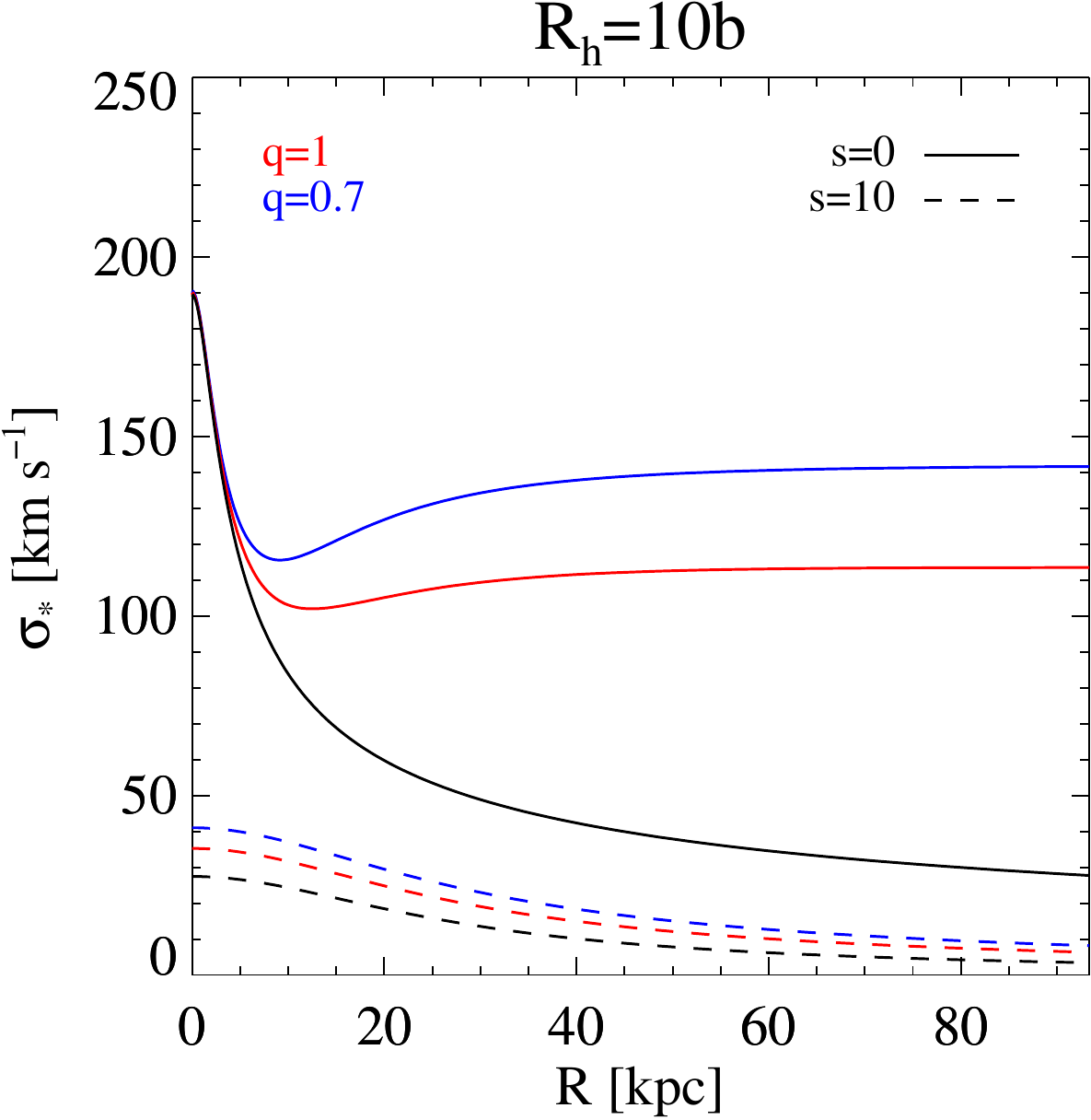}
\hspace{2mm}
\includegraphics[width=0.32\linewidth]{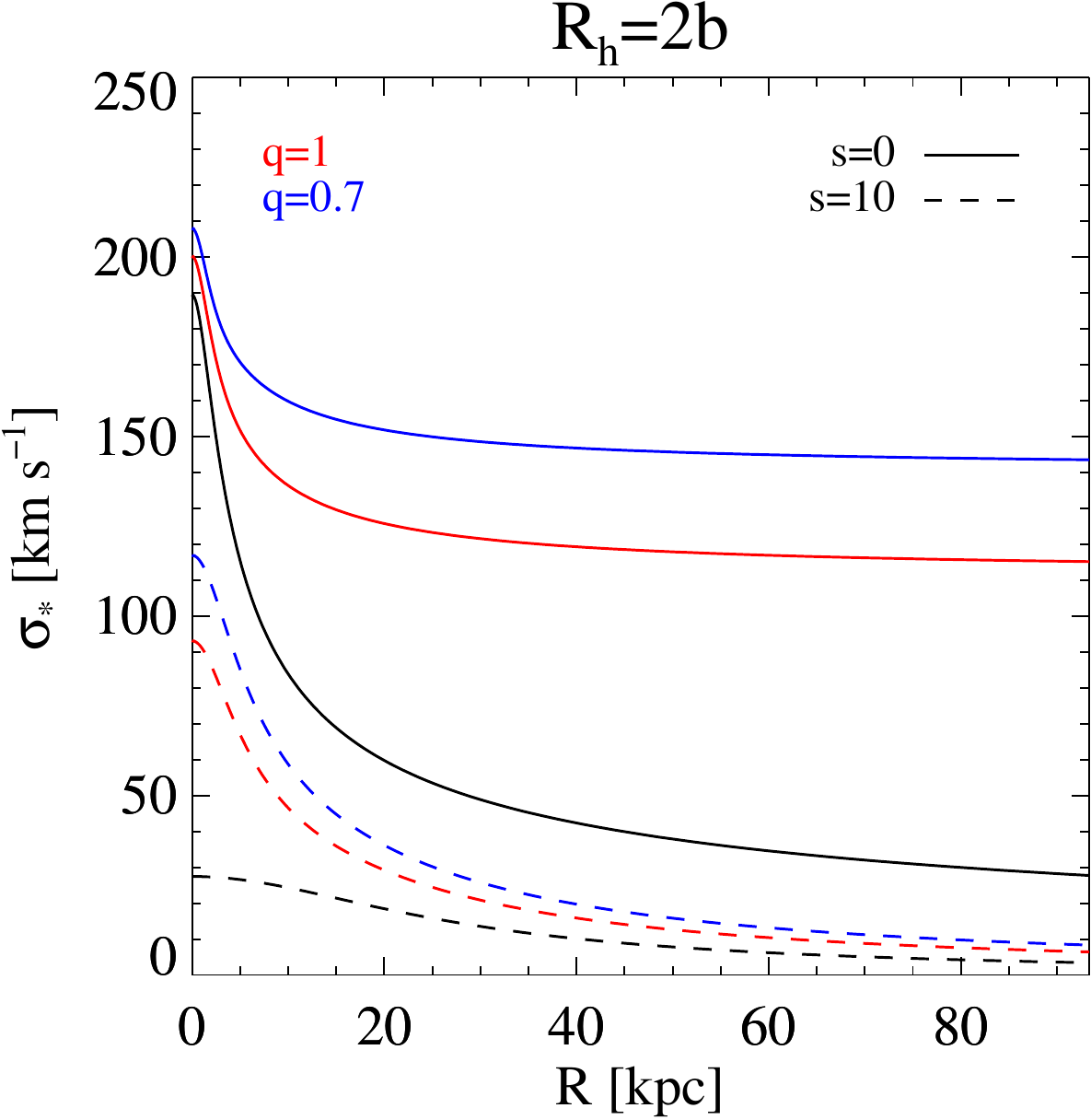}
\hspace{2mm}
\includegraphics[width=0.32\linewidth]{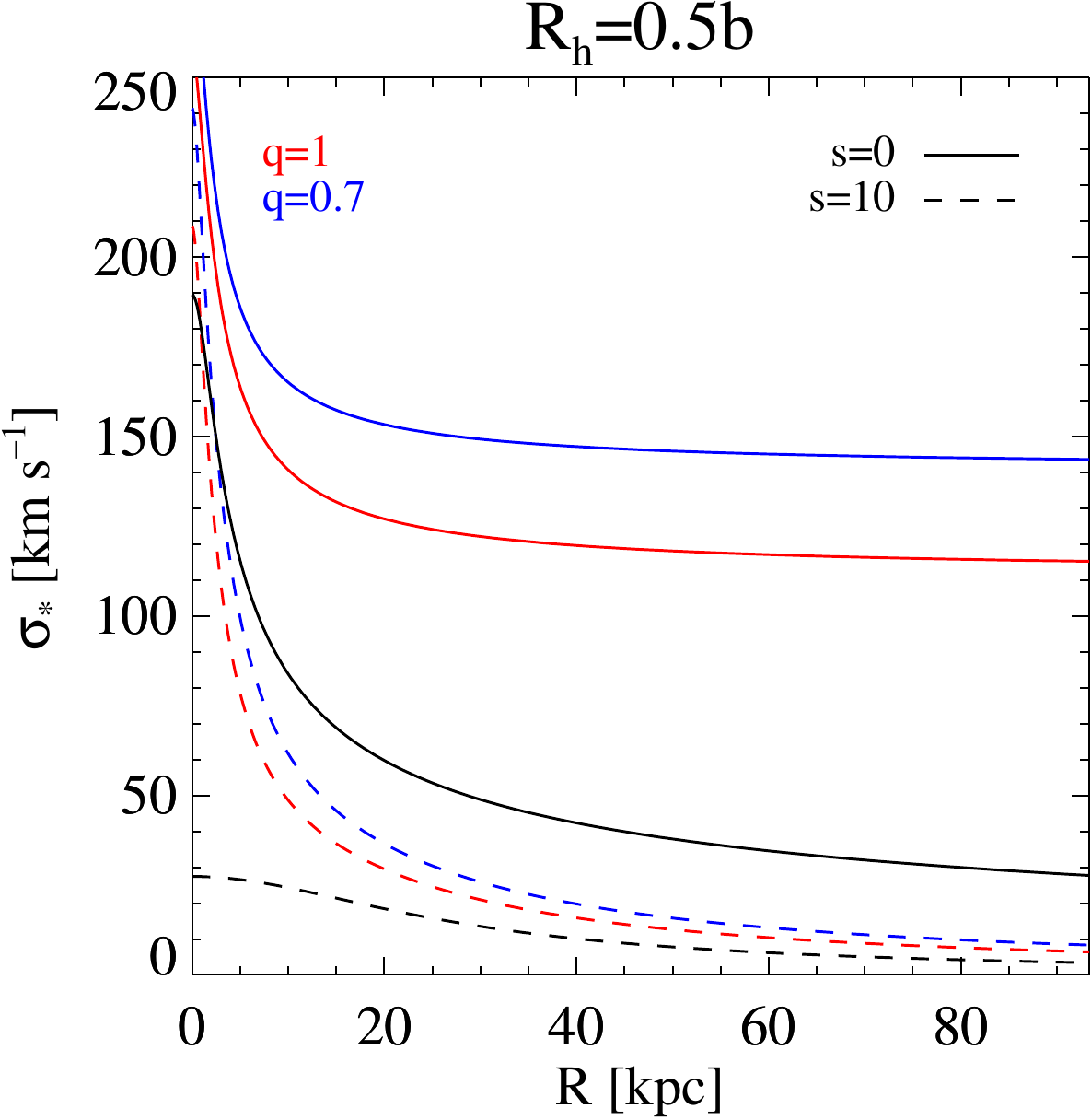}
\caption{Radial trend of the vertical velocity dispersion $\sigma_*$ in the equatorial plane for models with
$\Mstar=10^{11}\Msun$, $b=2$ kpc, $\vh=250\,\kms$, and different values of $s$, $q$, and $\Rh$. Solid
lines refer to a spherical stellar distribution ($s=0$), while the dashed lines to a flattened MN disc ($s=10$). Red and
blue lines correspond to spherical ($q=1$) and maximally flat ($q\simeq0.7$) equipotentials for the DM halo. For
reference, the black curves represent the stellar contribution $\sigma_{**}$. As discussed in the text, the models
represented by the blue solid lines cannot be described with the Satoh's ansatz, and in particular cannot be isotropic.}
\label{fig:sigma}
\end{figure*}

Additional information on the behaviour of the velocity dispersion field as a function of the various parameters of the
models can be obtained by inspection of Fig.~\ref{fig:sigma}, where we plot the radial trend of $\sigma_*^2$ in the
equatorial plane for a selection of models. In particular, the plots show how the flattening of the stellar distribution
and the dark matter halo shape and concentration affect the vertical velocity dispersion of the stellar component. For
reference, the black lines represent the contribution to the velocity dispersion due to the one-component MN model in
the spherical limit (solid, $s=0$) and for $s=10$ (dashed). Of course, the vertical velocity dispersion field is
independent of the amount of anisotropy in the azimuthal motions, so that these plots hold independently of the value of
$k$. We can notice a few, obvious features. First, the velocity dispersion of the model without halo is - for each model
- lower than the velocity dispersion in presence of the halo. Second, while the velocity dispersion declines at large
radii for the MN model, being dominated by the potential monopole term, it flattens to a constant value for models
embedded in the logarithmic halo, as expected from the dominance of its quasi-isothermal dark matter profile at large
radii. Third, note how a decrease of $\Rh$ at fixed $q$ and $\vh$ leads to an increase of the velocity dispersion, due
to the stronger concentration and gravitational field of the halo.
We stress again that at the present stage we are not attempting to reproduce any specific object; thus
the choice of the three values for $\Rh$ is just to illustrate the effects of a dark matter halo with a large, moderate,
and small scale length, at fixed asymptotic rotational velocity. Of course, ``cuspy'' haloes are expected to produce
dynamical features similar to the ones obtained for $\Rh=0.5b$. 

More interesting are the effects of the halo and stellar flattenings. In particular, for a fixed stellar distribution,
an increase of the halo flattening (at fixed $\Rh$
and $\vh$) \textit{increases} the stellar velocity dispersion at each radius (cf. the pairs of solid or dashed red and
blue lines in each panel). The velocity dispersion increase is a consequence of the increase of the vertical
gravitational field of the halo, that is more and more equatorially concentrated for decreasing $q$. However, an
increase of the flattening of the stellar density distribution (at fixed total stellar mass), for fixed halo, leads to a
\textit{decrease} of the stellar velocity dispersion (cf. solid and dashed lines of given colours). This can appear a
curious behaviour, as the same argument above about the intensity of the vertical gravitational field applies, but it is
not. In fact, it is true that the vertical gravitational field of the stellar distribution increases for increasing $s$,
but the stellar population now has - by construction - a shorter vertical scale-length, so its ``temperature'' decreases
accordingly: the two effects more than compensate, with a resulting net decrease of $\sigma_{*}$. This behaviour is by
no means a peculiarity of the present models, and it can be easily proved with simple algebra for the simpler family of
oblate Ferrers ellipsoids \citep{Ciotti.Lanzoni.1997}.

\begin{figure*}
\includegraphics[width=0.32\linewidth]{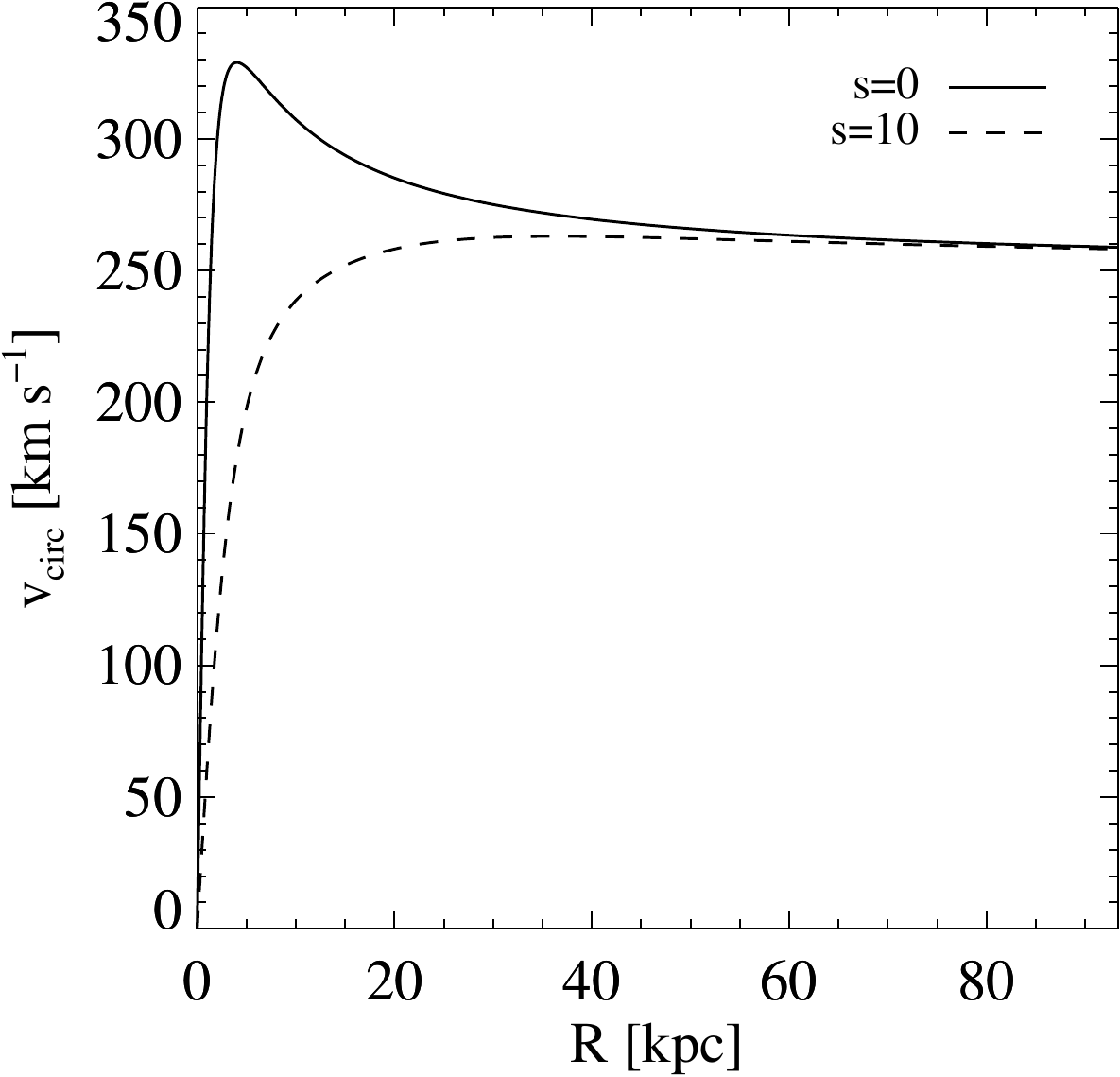}
\hspace{2mm}
\includegraphics[width=0.32\linewidth]{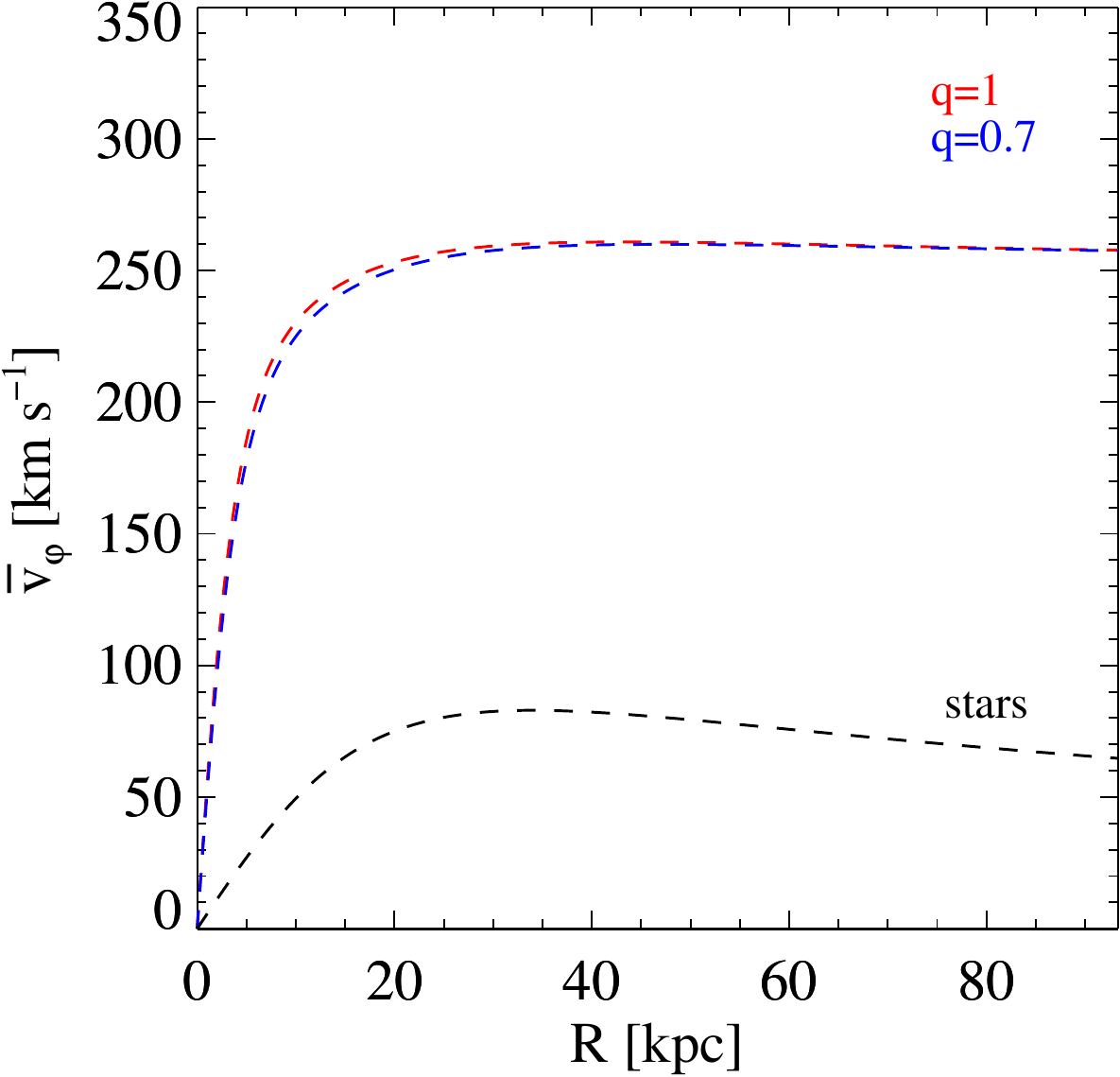}
\hspace{2mm}
\includegraphics[width=0.313\linewidth]{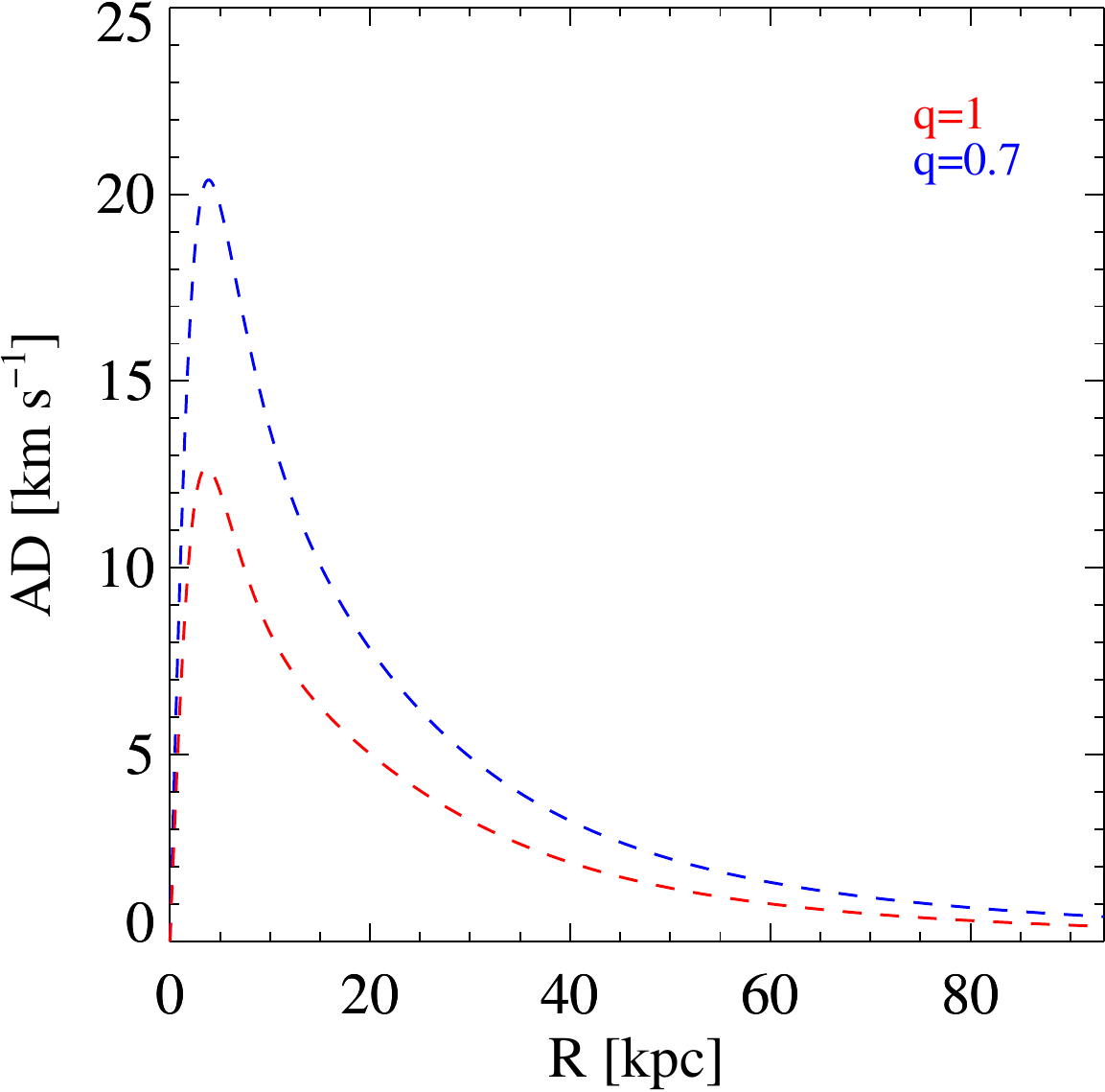}
\caption{
Dashed lines show the radial trend of the rotational properties of the stellar distribution in the
equatorial plane for the same models in the central panel of Fig.~\ref{fig:sigma} ($\Rh=2b$, $s=10$), and two
flattenings of the halo potential, $q=1$ (red lines) and $q=0.7$ (blue lines), in the isotropic case.
The solid line refers to a spherical stellar distribution ($s=0$). 
Left panel: the circular velocity $\vcirc$, a quantity independent of the halo flattening $q$ (see
eq.~\ref{eq:vcirc2}). Central panel: the streaming stellar velocity $\vphib$; for
reference, the black dashed curve represents the separate contribution of the stellar component.
Right panel: the Asymmetric Drift.}
\label{fig:appl}
\end{figure*}

Additional properties of the models are given in Fig.~\ref{fig:appl}, where we present, for the same models in the
central panel of Fig.~\ref{fig:sigma}, the rotational profiles of the stellar component in the equatorial plane, in the
isotropic rotator case. In particular, we show the circular velocity, the azimuthal streaming velocity,
and the Asymmetric Drift (defined as $\AD=\vcirc-\vphib$).
The most relevant features are the flattening of $\vphib$ to an asymptotic value that can be easily
calculated (see the following discussion). The streaming velocity is almost independent of the flattening of the halo
potential, a property reminiscent of $\vcirc$. Note that we do not show the rotational fields in the case of a spherical
density distribution. This is because for a spherical stellar density in a non-spherical halo, the Jeans equations in
the isotropic limit admit physical solution only for $q>1$, i.e., for a prolate halo potential. In fact, from
eq.~(\ref{eq:commutator}) it is immediate to show that for a generic spherical stellar density embedded in the
logarithmic potential in eq.~(\ref{eq:Binneypot}), the quantity $\overline{v_{\varphi}^2}-\sigma_{**}^2$ is proportional
to $q^2 -1$, and so the Satoh decomposition with $k>0$ cannot be adopted when $q<1$.

A more quantitative analysis of the effects of the model parameters on the dynamical properties
of the stellar population is provided by the asymptotic expansion of the solutions. The asymptotic expansion of the
dynamical quantities of the one-component MN model is trivial and can be obtained directly from
eqs.~(\ref{eq:jeans1solnohalo}) and (\ref{eq:jeans2solnohalo}), hence we do not discuss it here.
The asymptotic expansion of the formulae relative to the effects of the dark matter halo on the MN component can also be
obtained by working directly on the expressions in Section~\ref{sec:Binney}, however it
can also be obtained directly from the integral (\ref{eq:t-int}). Therefore, the following asymptotic trends (where
all lengths are normalized to $b$, and squared velocities are normalized to $G M_*/b$) have been obtained with both
methods, and have been also verified numerically with our Jeans solver code \citep{Posacki.etal2013},
thus giving an additional and independent check of the analytical integration.

We begin with the behaviour near the origin\footnote{It would be easy to consider the additional effect of a
supermassive central black hole on the velocity dispersion and on the circular velocity.} and we consider only the
expansion of $\sigma^2_*$, being the corresponding formulae for $\overline{\vphi^2}$ obtainable from
eq.~(\ref{eq:J2}). For a cored DM halo ($\Rh>0$) $\sigma_*^2$ and $\overline{\vphi^2}$ are finite at the origin, and
their value can be obtained directly from eq.~(\ref{IR0Ap}) and eq.~(\ref{eq:commutator}). As the resulting formulae are
trivial to obtain and not illuminating, we do not report them here. In the case $\Rh=0$, a coreless halo, $\sigma_*^2$
diverges at the centre due to the halo contribution as
\begin{equation}
\sigma_{*}^2=-\vh^2\ln\sqrt{z^2+q^2R^2}+\mathcal{O}(1),
\label{eq:asymporig}
\end{equation} 
where $\mathcal{O}(1)$ contains the term $\sigma^2_{**}(0)$ as obtained from eq.~(\ref{eq:jeans1solnohalo}).
These central trends can be proved also in a simpler way by considering the stellar component of a two-component model
that, in the central regions, behaves asymptotically as the MN model. This model is obtained by
superimposing to the Binney logarithmic halo the ellipsoidally stratified stellar density
\begin{equation}
\rho_*=\frac{\rho_{0*}R_{0*}^\beta}{(R_*^2+R^2+z^2/p^2)^{\beta/2}},
\label{eq:c1}
\end{equation}
where $R_{0*}$ and $R_*$ are two scale-lengths, $p$ is the axial ratio of the stellar isodensities, and $\beta>0$
controls the density slope outside the core. For $R_*=0$ the density is a pure power-law ellipsoid, while for $R_*>0$
the two scale-lengths can be assumed identical. In principle, the constants in eq.~(\ref{eq:c1}) could be fixed to match
the asymptotic expansion of the MN density distribution near the centre (in particular, $p^2=(5+s)/[(1+s)(5+4s+s^2)]$),
but this is not required by the following analysis. The vertical Jeans equation for this two-component model can be
integrated for generic positive values of $\beta$ in terms of the Incomplete Euler Beta function (i.e., hypergeometric
$\left._2F_1\right.$ functions, e.g., see eq.~8.391 in \citealt{Gradshteyn.etal2007}). Moreover, from
\citet{Chebyshev.1853} theorem on the integration of binomial differentials, the solution can be expressed in terms of
elementary functions for \textit{all rational values of $\beta$}. Additional elementary cases are obtained for $p=q$ and
$\Rh=R_*$ (e.g., see the $\beta=3$ case in \citealt{Binney.Tremaine.1987}). In the present case, it is easy to show that
$\sigma^2_*$ for $\rho_*$ in eq.~(\ref{eq:c1}), embedded in the potential in eq.~(\ref{eq:Binneypot}), diverges exactly
like eq.~(\ref{eq:asymporig}) for $\Rh=0$. Figure~\ref{fig:sigma} illustrates clearly the tendency of $\sigma_*$ to
diverge at the centre for $\Rh\to0$.

\begin{figure}
\centering
\includegraphics[width=0.7\linewidth]{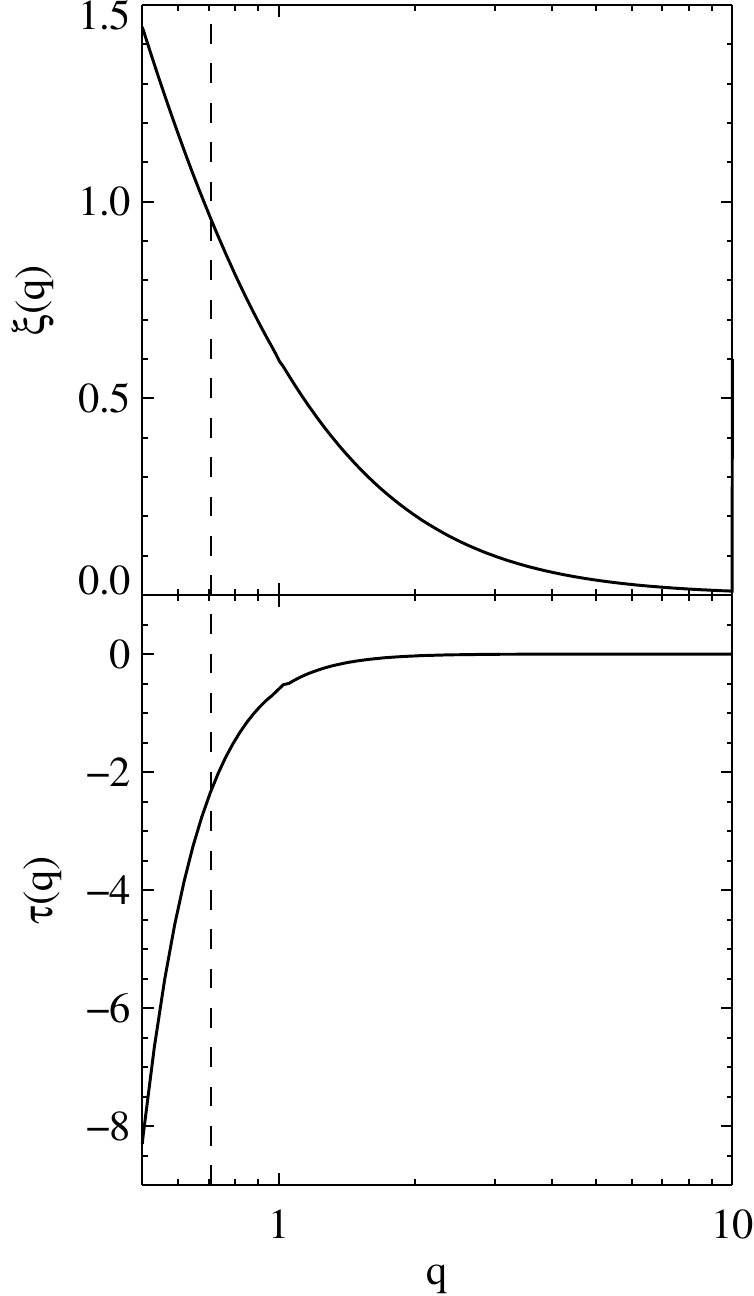}
\caption{The dimensionless functions $\xi(q)$ and $\tau(q)$ in the asymptotic expansion for $R\to\infty$ of the radial
and vertical velocity dispersion in the equatorial plane (eqs.~\ref{eq:csi} and \ref{eq:tau}).}
\label{fig:xitau}
\end{figure}

We now consider the asymptotic expansion at large distances from the origin. The full expression for $I$ along the $z$
axis ($R=0$) is given in Appendix~\ref{sec:app:R0}, and if we let $z$ tend to infinity it follows that
\begin{equation}
\sigma_{*}^2=\frac{1}{5}\vh^2+\mathcal{O}(z^{-1}),
\label{eq:asympaxis}
\end{equation}
so that the velocity dispersion is dominated by the contribution of the dark matter halo.
The treatment of the velocity dispersion in the equatorial plane ($z=0$) for $R\to\infty$ is considerably more
complicated. In particular, the asymptotic expansion of the integral~(\ref{eq:t-int}) can be obtained as a uniformly
convergent triple series of integer negative powers of $R$. A careful analysis shows that
\begin{equation}
\sigma_{*}^2=
\begin{cases}
\displaystyle \frac{\xi(q)}{3}\vh^2+\frac{1}{6R}+\mathcal{O}(R^{-2}), \\[2ex]
\displaystyle \left[\frac{1}{q^2}+\frac{\xi(q)}{s}\right]\frac{\vh^2}{R^2}+\left[\frac{(s+1)^2}{2s}+\vh^2\tau(q)\right]
\frac{1}{R^3} +\mathcal{O}(R^{-4}),
\end{cases}
\label{eq:asympeq}
\end{equation}
where the formulae above hold for $s=0$ and $s>0$, respectively. Note that the formula for the spherical case cannot be
obtained as the limit for $s\to0$ of the case with finite flattening of the stellar component, due to the
different behaviour of $\rhostar$ in eq.~(\ref{eq:rhostar}) for $a=0$ and $a>0$, respectively.
The two functions dependent on the halo flattening (see Fig.~\ref{fig:xitau}) are 
\begin{equation}
\xi(q) = \frac{q^2-4}{(q^2-1)^2}+\frac{3}{|q^2-1|^{5/2}}
\begin{cases}
\displaystyle \arctan\sqrt{q^2-1},& q>1, \\
\displaystyle\arctah\sqrt{1-q^2}, & q<1,
\end{cases} 
\label{eq:csi}
\end{equation}
with $\xi(1)=3/5$, and
\begin{equation}
\begin{split}
\tau(q)=&\frac{q^4-3q^2+17}{(q^2-1)^3}-\frac{1}{q^2}\\&+\frac{8q^4+8q^2-1}{q^3|q^2-1|^{7/2}}
\begin{cases}\displaystyle -\arctah\frac{\sqrt{q^2-1}}{q},& q>1, \\
\displaystyle\arctan\frac{\sqrt{1-q^2}}{q}, & q<1,
\end{cases}
\end{split}
\label{eq:tau}
\end{equation}
with $\tau(1)=-64/105$. All the trends in Fig.~\ref{fig:sigma} for $R\to\infty$ can be easily verified from the formulae
above.

In Fig.~\ref{fig:appl} we showed rotational properties of the stellar component in the equatorial plane, in the
isotropic case. Quantitatively, the circular velocity in the equatorial plane is 
\begin{equation}
v_{\rm circ}^2(R)=\frac{R^2}{\left[R^2+(s+1)^2\right]^{3/2}}+\frac{\vh^2R^2}{\Rh^2+R^2},
\label{eq:vcirc2}
\end{equation}
so that it is independent of the halo flattening $q$; for $R\to\infty$
\begin{equation}
v_{\rm circ}^2=\vh^2+\frac{1}{R}-\frac{\Rh^2\vh^2}{R^2}-\frac{3}{2}\frac{(s+1)^2}{R^3}+\mathcal{O}\left(R^{-4}\right),
\label{eq:vcircinfty}
\end{equation} 
while near the origin
\begin{equation} 
v_{\rm circ}^2=
\begin{cases}
\displaystyle\vh^2+\frac{R^2}{(s+1)^3}-\frac{3}{2}\frac{R^4}{(s+1)^5},\\[2ex]
\displaystyle\left[\frac{\vh^2}{\Rh^2}+\frac{1}{(s+1)^3}\right]R^2-\left[\frac{\vh^2}{\Rh^4}+\frac{3}{2(s+1)^5}\right]
R^4;
\end{cases}
\end{equation}
the formulae above are correct to $\mathcal{O}\left(R^6\right)$, and hold for $\Rh=0$ and $\Rh>0$, respectively.

A useful quantity related to the $\AD$ is given by the difference $\vcirc^2 -\vphib^2$, that in analytical studies is of
easier evaluation than the function $\vcirc -\vphib$. In fact, for moderate values of the Asymmetric Drift, it follows
that $\AD\simeq (\vcirc^2-\vphib^2)/(2\vcirc)$ \citep[e.g.,][]{Binney.Tremaine.1987}, while for example in the isotropic
case 
\begin{equation} 
\vcirc^2-\vphib^2=-\frac{R}{\rhostar}\frac{\partial \rhostar\sigma^2}{\partial R},
\end{equation}
as follows from eqs.~(\ref{eq:J2}) and (\ref{eq:Satohcons}). Using the asymptotic formula (\ref{eq:asympeq}), one
can show that
\begin{equation}
\vcirc^2-\vphib^2=
\begin{cases}
\displaystyle\frac{5}{3}\xi(q)\vh^2+\frac{1}{R}+\mathcal{O}(R^{-2}), \\[2ex]
\displaystyle\frac{5\vh^2}{R^2}\left[\frac{1}{q^2}+\frac{\xi(q)}{s}\right]+\frac{6}{R^3}\left[\frac{(s+1)^2}{2s}
+\vh^2\tau(q)\right]+\mathcal{O}(R^{-4}),
\end{cases}
\label{eq:delta}
\end{equation}
where the formulae hold for $s=0$ and $s>0$, respectively.
Notice that for fixed $s$, the leading order term of the $\vcirc^2-\vphib^2$ expansion is a decreasing function of $q$.
From eqs.~(\ref{eq:vcircinfty}) and (\ref{eq:delta}) one obtains the asymptotic formula for $\vphib$ in the equatorial
plane for the isotropic rotator, and at the leading order $\vphib\simeq\vh$ for all models with $s>0$.

We conclude presenting a qualitative consideration about gas flows in galactic discs, obtained by using one of the
results of this paper. In general it is expected (e.g., by chemical evolution studies) that disc galaxies host
\textit{radial} flows in their equatorial regions \citep[e.g.,][and references therein]{Spitoni.etal2014}.
These flows
influence the chemical and age gradients of stellar populations, with important consequences for galaxy formation and
evolution. The problem is \textit{how} such radial flows are sustained, i.e., what are the mechanisms responsible for
the redistribution of the angular momentum of the disc ISM. Proposed mechanisms range from loss of axisymmetry of the
gravitational field (e.g., bars, spiral arms, Q-instabilities, etc.), to viscosity (e.g., due to MRI), to gas
accretion on the disc of extra-planar material with low specific angular momentum
\citep[e.g.,][]{Lacey.Fall.1985,Elmegreen.etal2014}, so that after the mixing the gas falls toward the centre (as the
angular momentum of circular orbits, $J_0$, increases for increasing $R$ in stable discs). Here we consider an
additional, less stochastic mechanism that can lead to radial gas flows, related to internal phenomena, i.e., the
coupling between the stellar mass losses of the stars near the equatorial plane and the pre-existing gas. Of course, we
are not proposing that this is the \textit{only} mechanism at work, as accretion of external gas in star-forming disc
galaxies is required by pure mass budget arguments. However, it is interesting to consider the possibility that internal
phenomena may contribute to radial flows (see also \citealt{Bilitewski.Schonrich2012}).

\begin{figure}
\centering
\includegraphics[width=0.66\linewidth]{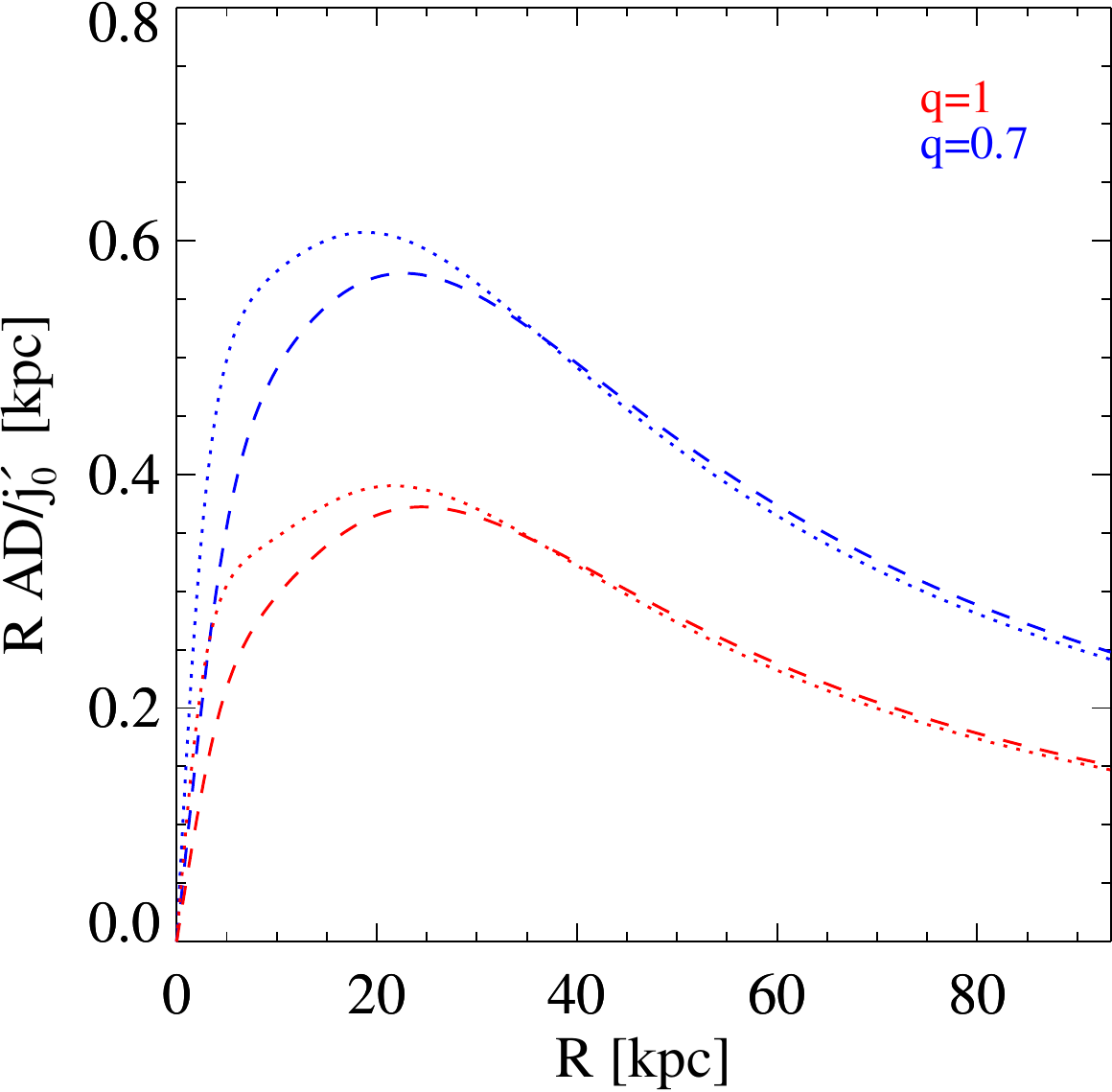}
\caption{Dashed lines show the radial trend in the equatorial plane of the function $R\times\AD/j'_0$,
appearing in eq.~(\ref{eq:vin}), for the same models in the central panel of Fig.~\ref{fig:sigma} ($\Rh=2b$, $s=10$),
and two flattenings of the halo potential, $q=1$ (red lines) and $q=0.7$ (blue lines), in the isotropic case. The dotted
lines show the quantity $R\times \AD/\vcirc$, which can be used as a proxy for the true
function (see the text below eq.~\ref{eq:vin} for additional explanations).}
\label{fig:radin}
\end{figure}

We derive the relevant relation between the $\AD$ and $\vin$, the expected inflow velocity of the gas towards the galaxy
central regions, restricting to the equatorial plane for the sake of simplicity. Consider some cold gas in the
annulus between $R$ and $R+\Delta R$, rotating with $\vcirc(R)$.
Then some other gas, coming from the evolving stellar population (e.g., \citealt{Ciotti.etal1991}), is added
with velocity $\vphib$, the streaming rotational velocity of the stars (e.g., see \citealt{DErcole.etal2000,
Negri.etal2013} for the hydrodynamical aspects). Because of the asymmetric drift, this new gas has a lower ordered
injection velocity than the
pre-existing gas, and hence a lower specific angular momentum. The two components mix and the resulting specific angular
momentum is the mass-weighted average of both specific angular momenta. The net result will be that the mixed gas will
fall inward with some inflow velocity $\vin$. Suppose the surface density of the cold gas
is given by $\Sigma(R)$, so that its angular momentum in the radial annulus between $R$ and $R + \Delta R$ is given by
\begin{equation} 
J_0(R)=2\pi R^2 \Delta R\,\Sigma(R)\, \vcirc(R).
\end{equation}
In a time interval $\delta t$ the evolving stars inject new material at a rate $\dot{\Sigma}(R)$, for an amount of mass 
\begin{equation} 
\Delta M_{\rm inj}=2\pi R\,\Delta R\,\dot{\Sigma}(R)\,\delta t,
\end{equation} 
and angular momentum
\begin{equation} 
\Delta J_{\rm inj}=2\pi R^2 \Delta R\, \dot{\Sigma}(R)\,\delta t \,\vphib.
\end{equation}
The angular momentum per unit time after the mixing of the new material with the pre-existing one is given by 
\begin{equation}
j(R)=\frac{R(\Sigma\,\vcirc+\dot{\Sigma}\,\delta t\, \vphib)}{\Sigma+\dot{\Sigma}\,\delta t}
 \simeq j_0(R)-R\,\AD\,\delta t\times \frac{\dot{\Sigma}}{\Sigma},
\end{equation} 
where $j_0(R)=R\vcirc(R)$ is the specific angular momentum of the
cold gas before injection.
Since $\AD>0$, there is a net radial inflow of gas to the radius $R+ \delta R$, defined by the condition $j_0(R+\delta
R) = j(R)$. Assuming a slow evolution (i.e., long characteristic times $\Sigma/\dot{\Sigma}$), and retaining linear
order terms in $\delta t$ and $\delta R$, one obtains an expression for the inflow velocity as
\begin{equation}
\vin(R)=- \dfrac{R\,\AD(R)}{j_0'(R)}\times\dfrac{\dot{\Sigma}(R)}{\Sigma(R)},\qquad\qquad
j_0'=\dfrac{\mathrm{d}j_0}{\mathrm{d}R},
\label{eq:vin}
\end{equation}
(see also eq.~3 in \citealt{Elmegreen.etal2014}).
Of course, a quantitative estimate of $\vin$ is beyond the scope of this paper, depending on the detailed temporal and
spatial distribution of $\dot{\Sigma}$ and $\Sigma$; however its order of magnitude can be
obtained for some typical case. For example, the dashed lines in Fig.~\ref{fig:radin} represent the
radial trend of $R\AD/j_0'$ for two specific models.
For a typical value of $\dot{\Sigma}/\Sigma=10^{-9} yr^{-1}$, these values lead to a velocity of the order of
$\vin\simeq1\,\kms$ around 20 kpc, consistent with the estimates required by chemical evolution models.
We notice that the exact term in eq.~(\ref{eq:vin}) in the case of a circular velocity described by a power law,
$\vcirc\propto R^\alpha$, would predict $R\AD/j'_0=R\AD/[(1+\alpha)\vcirc]$. This quantity (without the coefficient
$1+\alpha$) is shown in Fig.~\ref{fig:radin} with the dotted lines.

\section{Discussion and conclusions}\label{sec:conc}

In this paper we considered the two-integrals Jeans equations for a new family of two-component galaxy models described
by a Miyamoto-Nagai stellar disc embedded in a dark matter Binney logarithmic halo, a cored generalisation of the
singular isothermal sphere. We analytically solved these equations and gave the asymptotic behaviour of the quantities
involved, both at infinity and near the origin.
Moreover, we show that the solution of the Jeans equations for this model can also be obtained with the Residue
Theorem.

The obtained formulae have been tested against their numerical solutions, computed with our axisymmetric Jeans solver
code \citep{Posacki.etal2013}. Relative errors smaller than $\simeq 10^{-3}$ (or better) are found over the whole
computational domains for all the explored combinations of the model parameters. In turns, this is an independent check
of our Jeans solver code.

As a simple illustration of our model, we presented the effects of the stellar and dark matter halo flattenings on the
vertical velocity dispersion of stars in the equatorial plane. We showed that the velocity dispersion increases with
the flattening of the halo, and decreases for increasing flattening of the stellar distribution. As is well known, this
is relevant for studies of dark matter densities in the solar neighbourhood \citep[see,
e.g.,][]{King.Gilmore.vanderKruit1990,Binney.Tremaine.1987}.
We also used the asymmetric drift to obtain estimates for a possible new mechanism that could contribute to sustain
radial gas flows in disc galaxies, due to mass loss of stars near the equatorial plane.

The applications just mentioned and explored in a very preliminary way are just a few of many other possible
applications, e.g. the study of the circular velocity of gas in the equatorial plane, the building of hydrostatic,
barotropic and baroclinic models for hot rotating models \citep{Barnabe.etal2006}, the setup of numerical simulations of
gas flows in early-type galaxies with a proper description of thermalisation and stellar motions, etc. Our new model
provides a significant addition to the class of known fully analytical axisymmetric models with dark matter haloes. 

We notice that recently two papers \citep{Evans.Bowden.2014,Evans.Williams.2014} presented new analytical families of
axisymmetric dark matter haloes, based on interesting modifications of the Miyamoto-Nagai density-potential pair.
In fact, we checked that the Jeans equations for a MN model coupled to these new haloes can also be
solved analytically by using the approach here presented.

\section*{Acknowledgements}

We acknowledge the anonymous referee for useful comments, and Tim De Zeeuw and Wyn Evans for additional
comments.
This material is based upon work of L.C. supported in part by the National Science Foundation under Grant No. 1066293
and the hospitality of the Aspen Center for Physics. We acknowledge financial support from PRIN MIUR 2010-2011, project
``The Chemical and Dynamical Evolution of the Milky Way and Local Group Galaxies'', prot. 2010LY5N2T.

\bibliographystyle{mn2e}
\bibliography{references}{}

\appendix

\section[]{Integration procedure} \label{app:intcos}

We summarize the main steps needed to obtain the final formulae in
Section~\ref{sec:Binney}. The starting point is eq.~(\ref{eq:totalint}). The
special cases $R=0$ ($z$ axis), $A=0$ (critical cylinder), and $s=0$
(spherical stellar density) are treated in
Appendix~\ref{app:special}; the corresponding formulae for these special cases can be obtained as
limits of the general solution, but for the sake of simplicity we give
them explicitly.

A partial fraction decomposition of the integrand in eq.~(\ref{eq:totalint}) proves eq.~(\ref{eq:I123}): 
\begin{eqnarray}
\nonumber I&=&
\frac{1}{R^2}\int_{\arcsh\lambda}^\infty\frac{\alpha_0+\alpha_1\sh x+\alpha_2\sh^2x+\alpha_3\sh^3x}{(1+\sh^2x)^2}dx\\
&&\nonumber
+\frac{1}{R^2}\int_{\arcsh\lambda}^\infty\frac{\beta_0+\beta_1\sh
 x}{(R\sh x-s)^2}dx\\
&&\nonumber
+\frac{1}{R^2}\int_{\arcsh\lambda}^\infty\frac{\gamma_0+\gamma_1\sh x}{A+(R\sh x-s)^2}dx\\
&=&\frac{\Ia+\Ib+\Ic}{R^2},
\label{eq:IaIbIc}
\end{eqnarray}
where $\lambda$ is defined in eq.~(\ref{adef}), and the meaning of $\Ia, \Ib$ and $\Ic$ is obvious.

\subsection{The integral $\Ia$}\label{sec:app:Ia}

The partial fraction decomposition coefficients of $\Ia$ in
eq.~(\ref{eq:IaIbIc}) for $R\neq 0$, $A\neq 0$, and $s\neq 0$ are given by
\begin{equation}
\begin{split}
\dfrac{\alpha_0\alpha_d}{s}=&-17R^8-2R^6(11s^2-12A)\\&+R^4(8s^4+19s^2A-9A^2)\\
&+2R^2(s^2+A)(7s^4+A^2)+s^2(s^2+A)^3,\\
\dfrac{\alpha_1\alpha_d}{R}=&-6R^8+R^6(14s^2+15A)\\&+2R^4(25s^4-4s^2A-6A^2)\\
&+R^2(34s^6+3s^4A+4s^2A^2+3A^3)\\
&+2s^2(2s^6+5s^4A+4s^2A^2+A^3),\\
\dfrac{\alpha_2\alpha_d}{s}=&-8R^8+R^6(2s^2+3A)+2R^4(13s^4+8s^2A+3A^2)\\
&+R^2(s^2+A)(14s^4+9s^2A-A^2)-2s^2(s^2+A)^3,\\
\dfrac{\alpha_3\alpha_d}{R}=&-3R^8+2R^6(7s^2+3A)\\&+R^4(32s^4+s^2A-3A^2)+2s^2R^2(5s^4-A^2)\\
&-s^2(5s^2+A)(s^2+A)^2,
\end{split}
\end{equation}
where
\begin{equation}
\alpha_d=(R^2+s^2)^2[(A+s^2-R^2)^2+4R^2s^2]^2.
\end{equation}
The integration of the various terms is elementary.

\subsection{The integral $\Ib$}\label{sec:app:Ib}

The partial fraction decomposition coefficients of $\Ib$ in eq.~(\ref{eq:IaIbIc}) for $R\neq 0$, $A\neq 0$, and $s\neq
0$ are given by
\begin{equation}
\beta_0\beta_d=R^4s, \qquad \beta_1\beta_d=R^3s^2,
\end{equation}
where 
\begin{equation}
\beta_d=A(R^2+s^2)^2.
\end{equation} 
Using the standard substitution $y=\tah(x/2)$ we obtain
\begin{equation}
I_\beta=-\frac{2\beta_0}{s^2}\int_{\mu}^1\frac{y^2-2sy/R-1}{(y^2+2Ry/s-1)^2}dy,
\end{equation} 
where $\mu$ is given by eq.~(\ref{eq:y0}). It can be easily proved
that the two real zeros of the denominator lie outside the integration
domain. Elementary integration simplifies to the
surprisingly simple expression in eq.~(\ref{eq:Ibsol}).

\subsection{The integral $\Ic$}\label{sec:app:Ic}

The partial fraction decomposition coefficients of $\Ic$ in eq.~(\ref{eq:IaIbIc}) for $R\neq 0$, $A\neq 0$, and $s\neq
0$ are given by

\begin{equation}
\begin{split}
\frac{\gamma_0\gamma_d}{R^2s}=&-R^6-2R^4(s^2-3A)-R^2(s^4-18s^2A-3A^2)\\
&-8A(s^2+A)^2,\\
\frac{\gamma_1\gamma_d}{R^3}=&-R^4(s^2-3A)-2R^2(s^4+8s^2A+3A^2)\\
&-(s^2-3A)(s^2+A)^2,
\end{split}
\end{equation}
where
\begin{equation} 
\gamma_d=A[(A+s^2-R^2)^2+4R^2s^2]^2.
\end{equation}

At variance with the integrals $\Ia$ and $\Ib$, the integration procedure now depends on the sign of $A$. Inspection of
eq.~(\ref{eq:IaIbIc}) suggests that an easy factorisation of the denominator of $\Ic$ could be obtained in the case
$A<0$. However, as the same procedure cannot be applied to the case $A>0$ without using complex numbers, we prefer to
follow another approach that maximises the similarity of the treatment in the two cases. 

The substitution $y=e^x$ in the last integral in eq.~(\ref{eq:IaIbIc}) gives
\begin{equation}
I_\gamma=\frac{2\gamma_1}{R^2}\int_\nu^\infty\frac{y^2+Hy-1}{\Delta(y)}\:dy,
\end{equation}
where 
\begin{equation} 
H=2\frac{\gamma_0}{\gamma_1},
\end{equation}
\begin{equation}
\Delta(y)=y^4-\frac{4s}{R}y^3+\left(\frac{4A}{R^2}+\frac{4s^2}{R^2}-2\right)y^2+\frac{4s}{R}y+1,
\label{eq:Delta}
\end{equation}
and $\nu$ is given by eq.~(\ref{eq:nu}).

In principle, we could use the antisymmetry of $\Delta(y)/y^2$ (after noticing that $y=0$ is not a zero) to factorise
it: it is readily seen that if $y_1$ is a zero of $\Delta$, then so is $-1/y_1$. This implies that $\Delta(y)/y^2$ can
be written as a quadratic polynomial in $t=y-1/y$, from which the factorisation is immediate. However, if $A>0$ the two
roots of the quadratic polynomial in $t$ are complex conjugates.

In practice this computation is not needed since any quartic polynomial with real coefficients can be factorised into
two quadratic polynomials with real coefficients. Without loss of generality, we found it useful to adopt the
factorisation

\begin{equation}
\Delta(y)=\left[(y-\Delta_+)^2+\delta_+\right]\left[(y-\Delta_-)^2+\delta_-\right].
\label{eq:Deltafac}
\end{equation} 
Expansion of eq.~(\ref{eq:Deltafac}) and comparison with eq.~(\ref{eq:Delta}) shows that 
\begin{equation}
\Delta_{\pm}=\frac{s}{R}\pm\sqrt{\frac{s^2}{R^2}+\delta},\qquad
\delta_\pm=\frac{1-\delta}{\delta}\Delta_{\pm}^2,\\
\end{equation}
where 
\begin{equation} 
\delta=\frac{\sqrt{(A+s^2-R^2)^2+4R^2s^2}-(A+s^2-R^2)}{2R^2}.
\end{equation}

Notice that $\delta>0, \Delta_+>0, \Delta_-<0$. If $A>0$, then $0<\delta<1$ and hence $\delta_{\pm}>0$, making the two
quadratic polynomials in eq.~(\ref{eq:Deltafac}) irreducible over the reals. If $A<0$, then $\delta>1$ and hence
$\delta_{\pm}<0$, consistent with the fact that in this case $\Delta$ can be factorised into four linear factors over
the reals.

Now we can proceed in the usual way, by a partial fraction decomposition. The coefficients in 
\begin{equation}
\frac{y^2+Hy-1}{\Delta(y)}=\frac{\eta_+y+\theta_+}{(y-\Delta_+)^2+\delta_+}+\frac{\eta_-y+\theta_-}{
(y-\Delta_-)^2+\delta_-}
\end{equation} 
are given, after some simplification, by 
\begin{equation}
\begin{split}
\eta_\pm\sigma_d=&\pm 2\delta\left(2\delta-\frac{Hs}{R}\right),\\
\theta_\pm\sigma_d=&2\Delta_\pm\left[\frac{s}{R}(\Delta_+-\Delta_-)\pm\delta\Delta_\pm(H+2\Delta_\mp)\right],
\end{split}
\end{equation}
where 
\begin{equation}
\sigma_d=4(\Delta_+-\Delta_-)\left(\delta^2+\frac{s^2}{R^2}\right).
\end{equation} 
Note that a few simple algebraic relations, useful to simplify the final expression of $\Ic$ in eq.~(\ref{eq:Icsol}),
link the constants above: 
\begin{equation}
\Delta_+^2\delta_-=\delta_+\Delta_-^2\:,\qquad
\frac{\eta_+\Delta_++\theta_+}{\Delta_+}=\frac{\eta_-\Delta_-+\theta_-}{\Delta_-}.
\end{equation} 

\section[]{Special cases} \label{app:special}

In the following we give the explicit solution of eq.~(\ref{eq:totalint}) in the special cases
$R=0$, $A=0$, $s=0$, when the formulae in Section~\ref{sec:Binney} cannot be used.
We recall that $A=q^2(R^2+\Rh^2)-1$.

\subsection{Velocity dispersion on the $z$ axis}\label{sec:app:R0}

For $R=0$ (i.e., along the $z$-axis), eq.~(\ref{eq:t-int}) simplifies considerably, and its
integration is elementary:
\begin{equation}
\begin{split}
I=&\frac{1}{As^2\zeta}+\frac{4s}{(A+s^2)^3}\ln\frac{(\zeta+s)^2}{\zeta^2+A}-\frac{A+5s^2}{s^2(A+s^2)^2(\zeta+s)}
\\&-\frac{1}{s(A+s^2)(\zeta+s)^2}\\
&+\frac{3A^2-6As^2-s^4}{A\sqrt{|A|}(A+s^2)^3}
\begin{cases}
\displaystyle\arctan\frac{\sqrt{A}}{\zeta},& A>0,\\
\displaystyle\arctah\frac{\sqrt{|A|}}{\zeta},& -1\leq A<0.
\end{cases}
\end{split}
\label{IR0Ap}
\end{equation} 

The cases $A=0$ and $A=-s^2$ should be treated separately. For $A=0$, i.e., when the critical cylinder coincides with
the $z$-axis ($\Rh=1/q$), then 
\begin{equation}
I=\frac{8}{s^5}\ln\frac{\zeta+s}{\zeta}+\frac{s^4+2s^3\zeta-8s^2\zeta^2-36s\zeta^3-24\zeta^4}{3s^4\zeta^3(\zeta+s)^2}.
\label{IR0A0}
\end{equation}
The case $A=-s^2$ is possible only for $s<1$, so that $\zeta >s$, and 
\begin{equation}
I=\frac{1}{4s^5}\ln\frac{\zeta+s}{\zeta-s}-\frac{6s^3+10s^2\zeta+9s\zeta^2+3\zeta^3}{6s^4\zeta(\zeta+s)^3}.
\label{IR0Anse}
\end{equation} 

Notice that this solution is always finite, except at the origin ($z=0$, i.e. $\zeta=1$) for $s=1$, so that $A=-1$ on
the $z$-axis, which in turn implies $\Rh=0$, i.e., when the dark matter potential is not cored.

\subsection{Velocity dispersion on the critical cylinder}\label{sec:I2A0}

On the critical cylinder $R^2=\Rc^2\equiv q^{-2}-\Rh^2$ the parameter
$A$ vanishes, and two denominator factors in eq.~(\ref{eq:totalint})
coincide. Note that, if $q\Rh>1$, there is no critical cylinder since
$A>0$ for every $R$. If $q\Rh<1$ then $\Rc$ exists, in particular $\Rc=1$
for the SIS model. If $q\Rh=1$ then the critical cylinder coincides
with the $z$ axis and the solution for $I$ is given by
eq.~(\ref{IR0A0}).

The partial fraction decomposition in eq.~(\ref{eq:IaIbIc}) is no longer valid, instead we have that 
\begin{equation}
I=\frac{I_{\alpha}+I_c}{\Rc^2},
\label{eq:Icrit} 
\end{equation}
where $I_\alpha$ is as before and $I_c$ can, without loss of generality, be written as 
\begin{equation}
I_c=\sum_{i=1}^4\int_{\arcsh\lambda}^\infty\frac{\theta_i}{(\Rc\sh x-s)^i}dx,
\label{eq:Icritdef}
\end{equation}
where no singularities are contained in the integration domain, and the coefficients $\theta_i$ can be found by the
usual partial fraction decomposition technique. 
The substitution $y=e^x$ transforms the integrals in rational ones, and in particular
\begin{equation}
\begin{split}
\int\frac{dx}{\Rc\sh x-s}&=\frac{2}{\Rc}\int\frac{dy}{y^2-2\frac{s}{\Rc}y-1}\\
&=-\frac{2}{\sqrt{\Rc^2+s^2}}\arctanh\frac{\sqrt{\Rc^2+s^2}}{\Rc-s}.
\end{split}
\label{eq:b6}
\end{equation} 
The other integrals for $i=2,3,$ and 4 are most easily obtained by differentiating with respect to $s$
eq.~(\ref{eq:b6}). The limits of integration for $y$ are $\nu_c$ given by eq.~(\ref{eq:nu}) evaluated at
$R=\Rc$, and $\infty$. The final result for $I_c$ is
\begin{equation}
\begin{split}
I_c=&\frac{2\Rc^2(3\Rc^4-24\Rc^2s^2+8s^4)}{(\Rc^2+s^2)^{9/2}}\arctanh\frac{\sqrt{\Rc^2+s^2}}{\Rc\nu_c-s}\\
&+\frac{2}{3}\frac{\Rc^2s}{(\Rc^2+s^2)^4}\frac{P_5(\nu_c,\Rc)}{(\Rc\nu_c^2-2s\nu_c-\Rc)^3},
\end{split}
\end{equation}
where 
\begin{equation}
\begin{split}
P_5(\nu,R)=&3R^2s(4R^2-3s^2)\nu^5\\&+R(15R^4-54R^2s^2+36s^4)\nu^4\\
&+s(-78R^4+100R^2s^2-32s^4)\nu^3\\&+6R(-4R^4+21R^2s^2-10s^4)\nu^2\\
&+3R^2s(22R^2-13s^2)\nu+R^3(13R^2-8s^2).
\end{split}
\end{equation}
A careful treatment shows that in the special case where the critical cylinder coincides with the $z$ axis (i.e.
$q\Rh=1$ and $A=R=0$), eq.~(\ref{eq:Icrit}) coincides with eq.~(\ref{IR0A0}).

\subsection{Spherical stellar density}\label{sec:app:plummer}
In the case of a spherical stellar density, i.e., when the MN model reduces to the Plummer sphere, eq.~(\ref{eq:t-int})
simplifies to 
\begin{equation} 
I=3\int_\zeta^\infty\frac{\zeta'}{(R^2+\zeta'^2)^{5/2}(A+\zeta'^2)}d\zeta'.
\end{equation} 
The substitution $u=\sqrt{\zeta^2+R^2}$ gives a rational integrand, and
\begin{equation}
\begin{split}
I=&\frac{1}{(A-R^2)(R^2+\zeta^2)^{3/2}}-\frac{3}{(A-R^2)^2\sqrt{R^2+\zeta^2}}\\
&+\frac{3}{|A-R^2|^{5/2}}
\begin{cases}
\displaystyle\arctan\sqrt{\frac{A-R^2}{R^2+\zeta^2}},&A>R^2,\\
\displaystyle\arctah\sqrt{\frac{R^2-A}{R^2+\zeta^2}},&A<R^2,
\end{cases}
\end{split}
\label{eq:sphgen}
\end{equation}
while for $A=R^2$
\begin{equation}
I=\frac{3}{5(R^2+\zeta^2)^{5/2}}.
\label{eq:sphspec}
\end{equation}

\label{lastpage}

\end{document}